\documentclass[longauth]{aa}  
\usepackage[dvipsnames]{xcolor}
\usepackage{graphicx}
\usepackage{amsmath}
\usepackage{csvsimple}
\usepackage{rotating}
\usepackage{longtable}
\usepackage{float}
\usepackage{tikz}
\usepackage[locale = UK, detect-all]{siunitx}
\usepackage{subcaption}
\usepackage{lscape}
\usepackage{mathtools}
\usepackage{txfonts}
\usepackage{hyperref}
\hypersetup{colorlinks=true,citecolor=blue}

\newcommand{\HII}{H{\sc ii}}
\newcommand{\flowrate}{10$^{-4}$~M$_{\sun}\mathrm{yr}^{-1}$}

\begin{document} 
\title{Dynamical Accretion Flows}
\subtitle{ALMAGAL: Flows along filamentary structures in high-mass star-forming clusters}
\author{M. R. A. Wells\inst{1}\and H. Beuther\inst{1}\and S. Molinari\inst{2}\and P. Schilke\inst{3}\and C. Battersby\inst{4}\and P. Ho\inst{5,}\inst{6}\and Á. Sánchez-Monge\inst{7,8}\and B. Jones\inst{3}\and  M. B. Scheuck\inst{1}\and J. Syed\inst{1} \and C. Gieser\inst{38} \and R. Kuiper\inst{25} \and D. Elia\inst{2} \and A. Coletta\inst{2,24} \and A. Traficante\inst{2} \and J. Wallace\inst{4} \and A. J. Rigby\inst{39} \and R. S. Klessen\inst{9,10} \and Q. Zhang\inst{11} \and S. Walch\inst{3,12} \and M.\ T.\ Beltr\'an\inst{13} \and Y. Tang\inst{5} \and G. A. Fuller\inst{3,14} \and D. C. Lis\inst{15} \and T. Möller\inst{3} \and  F. van der Tak\inst{16,17} \and P. D. Klaassen\inst{18} \and  S. D. Clarke\inst{3,5} \and L. Moscadelli\inst{13} \and C. Mininni\inst{2} \and H. Zinnecker\inst{37} \and  Y. Maruccia\inst{2} \and S. Pezzuto\inst{2} \and M. Benedettini\inst{2} \and J. D. Soler\inst{2} \and C. L. Brogan\inst{19} \and A. Avison\inst{14,20,21} \and P. Sanhueza\inst{22,23} \and E. Schisano\inst{2} \and T. Liu\inst{26} \and F. Fontani\inst{13,27,28} \and K. L. J. Rygl\inst{29} \and F. Wyrowski\inst{30} \and J. Bally\inst{31} \and D. L. Walker\inst{20} \and A. Ahmadi\inst{32} P. Koch\inst{5} \and M. Merello\inst{33} \and C. Y. Law\inst{34,35} \and L. Testi\inst{13,36}}

\institute{Max-Planck-Institut f\"ur Astronomie, K\"onigstuhl 17, D-69117 Heidelberg, Germany\and INAF-Istituto di Astrofisica e Planetologia Spaziale, via Fosso del Cavaliere 100, 00133, Roma, Italy \and I. Physikalisches Institut, Universität zu K\"oln, Z\"ulpicher Str. 77, 50937 K\"oln, Germany \and University of Connecticut, Department of Physics, 2152 Hillside Road, Unit 3046 Storrs, CT 06269, USA \and Institute of Astronomy and Astrophysics, Academia Sinica, Taipei 10617, Taiwan \and East Asian Observatory, 660 N. A’oh\-ok\-u, Hilo, Hawaii, HI 96720, USA \and Institut de Ciéncies de l’Espai (ICE, CSIC), Can Magrans s/n, 08193, Bellaterra, Barcelona, Spain \and Institut d’Estudis Espacials de Catalunya (IEEC), Barcelona, Spain \and Universit\"at Heidelberg, Zentrum für Astronomie, Institut für Theoretische Astrophysik, Heidelberg, Germany \and Universit\"at Heidelberg, Interdisziplin\"ares Zentrum f\"ur Wissenschaftliches Rechnen, Heidelberg, Germany \and Harvard-Smithsonian Center for Astrophysics, 160 Garden St, Cambridge, MA 02420, USA \and Center for Data and Simulation Science, University of Cologne, Germany \and INAF – Osservatorio Astrofisico di Arcetri, Largo E. Fermi 5, 50125 Firenze, Italy \and Jodrell Bank Centre for Astrophysics \& UK ALMA Regional Centre Node, School of Physics \& Astronomy, University of Manchester, M13 9PL, UK \and Jet Propulsion Laboratory, California Institute of Technology 4800 Oak Grove Drive, Pasadena, CA, 91109, USA \and SRON Netherlands Institute for Space Research, Landleven 12, 9747 AD Groningen, The Netherlands \and Kapteyn Astronomical Intsitute, University of Groningen, The Netherlands \and UK Astronomy Technology Centre, Royal Observatory Edinburgh, Blackford Hill, Edinburgh EH9 3HJ, UK \and National Radio Astronomy Observatory (NRAO), 520 Edgemont Rd, Charlottesville, VA 22903, USA \and UK ALMA Regional Centre Node, M13 9PL, UK \and SKA Observatory, Jodrell Bank, Lower Withington, Macclesfield, SK11 9FT \and National Astronomical Observatory of Japan, National Institutes of Natural Sciences, 2-21-1 Osawa, Mitaka, Tokyo 181-8588, Japan \and Department of Astronomical Science, SOKENDAI (The Graduate University for Advanced Studies), 2-21-1 Osawa, Mitaka, Tokyo 181-8588, Japan \and Dipartimento di Fisica, Universita di Roma La Sapienza, Piazzale Aldo Moro 2, I-00185, Roma, Italy \and Faculty of Physics, University of Duisburg-Essen, Duisburg, Germany. \and Shanghai Astronomical Observatory, Chinese Academy of Sciences, 80 Nandan Road, Shanghai 200030, P. R. China \and Centre for Astrochemical Studies, Max-Planck-Institute for Extraterrestrial Physics, Giessenbachstrasse 1, 85748 Garching, Germany
\and LERMA, Observatoire de Paris, PSL Research University, CNRS, Sorbonne Universite, F-92190 Meudon (France) \and INAF-Istituto di Radioastronomia \& Italian ALMA Regional Centre, Via P. Gobetti 101, I-40129 Bologna, Italy \and Max-Planck-Institut fur Radioastronomie (MPIfR), Auf dem H\"ugel 69, 53121 Bonn, Germany \and Department of Astrophysical and Planetary Sciences, University of Colorado, Boulder, CO 80389, USA \and Leiden Observatory, Leiden University, PO Box 9513, 2300 RA Leiden, The Netherlands \and Departamento de Astronomía, Universidad de Chile, Santiago, Chile \and European Southern Observatory, Karl-Schwarzschild-Strasse 2, D-85748 Garching, Germany \and Department of Space, Earth \& Environment, Chalmers University of Technology, SE-412 96 Gothenburg, Sweden \and Dipartimento di Fisica e Astronomia ”Augusto Righi” Viale Berti Pichat 6/2, Bologna \and Universidad Autónoma de Chile, Nucleo Astroquimica y Astrofisica, Avda Pedro de Valdivia 425, Providencia, Santiago de Chile \and Max Planck Institute for Extraterrestrial Physics, Giessenbachstraße 1, 85749, Garching bei M\"unchen, Germany \and School of Physics and Astronomy, University of Leeds, Leeds LS2 9JT, UK}

\date{Received XXX; accepted XXX}

\abstract{Investigation of the flow of material along filamentary structures towards the central core can help provide insight into high-mass star formation and evolution.}{Our main motivation is to answer the question: what are the properties of accretion flows in star-forming clusters? We use data from the ALMA Evolutionary Study of High Mass Protocluster Formation in the Galaxy (ALMAGAL) survey to study 100 ALMAGAL regions at $\sim$ 1\arcsec~ resolution located between $\sim$ 2 and 6~kpc distance.}{Making use of the ALMAGAL $\sim$ 1.3mm line and continuum data we estimate flow rates onto individual cores. We focus specifically on flow rates along filamentary structures associated with these cores. Our primary analysis is centered around position velocity cuts in H$_2$CO (3$_{0,3}$ - 2$_{0,2}$) which allow us to measure the velocity fields, surrounding these cores. Combining this work with column density estimates we derive the flow rates along the extended filamentary structures associated with cores in these regions.}{We select a sample of 100 ALMAGAL regions covering four evolutionary stages from quiescent to protostellar, Young Stellar Objects (YSOs), and \HII~regions (25 each). Using dendrogram and line analysis, we identify a final sample of 182 cores in 87 regions. In this paper, we present 728 flow rates for our sample (4 per core), analysed in the context of evolutionary stage, distance from the core, and core mass. On average, for the whole sample, we derive flow rates on the order of $\sim$\flowrate~with estimated uncertainties of $\pm$50\%. We see increasing differences in the values among evolutionary stages, most notably between the less evolved (quiescent/protostellar) and more evolved (YSO/\HII~region) sources and we also see an increasing trend as we move further away from the centre of these cores. We also find a clear relationship between the calculated flow rates and core masses $\sim$M$^{2/3}$ which is in line with the result expected from the tidal-lobe accretion mechanism. The significance of these relationships is tested with Kolmogorov-Smirnov and Mann-Whitney U tests.}{Overall, we see increasing trends in the relationships between the flow rate and the three investigated parameters; evolutionary stage, distance from the core, and core mass.}

\maketitle

\section{Introduction}

The formation and evolution of high-mass stars has been the subject of intense scientific interest for decades. High-mass stars play a crucial role in shaping not only their parental clouds but also the interstellar medium on kpc scales, enriching it with heavy elements, and influencing the dynamics of their surrounding environments via the energy they release through radiation and stellar winds (e.g.,~\citealt{1974Kahn, 1977Yorke, 1987Wolfire, 2007ZY, 2007Arce, 2014Frank, 2009Smith, 2015Zhang, 2018Motte, 2018Kuiper}). This in turn triggers new waves of star formation and helps sculpt the physical conditions of the Interstellar Medium (ISM) in galactic disks. Therefore, understanding the intricate processes involved in the birth and subsequent evolution of high-mass stars is fundamental not only for stellar physics but also for comprehending the broader aspects of galaxy formation and evolution. High-mass stars are rare due to their short lifetimes and comparatively low numbers when compared to low-mass stars. Looking at the initial mass function (IMF) we see one reason they are in limited numbers is that the IMF at high mass values follows a power law (e.g,~\citealt{Salpeter,2007Bonnell, 2014Offner}). Moreover, high-mass stars typically stay embedded in their natal clusters until they reach the main sequence, making it much more difficult for us to study and constrain how they form and evolve. This leaves us with a large knowledge gap in not only star formation but astrophysics in general. 


What we do know, is that the most common place for star formation to occur is in clustered environments inside giant molecular clouds (GMCs) (e.g.,~\citealt{Lada_2003, 2010Bressert}). These are immense reservoirs of cold, dense gas and dust that provide the material for new generations of stars. Molecular clouds are commonly described to have a hierarchical structure (e.g,~\citealt{1985Scalo, 2022Thomasson}). Following \cite{2000Williams} and \cite{2007Beuther} these clouds host massive condensations of gas called clumps ($\sim$ 1~pc), which form clusters, within which more compact cores ($\sim$ 10000~AU) are observed that form gravitationally bound single, binary, or multiple systems. The process begins with the fragmentation of these GMCs due to gravitational instabilities, resulting in the formation of clumps and cores (e.g., \citealt{1984MZinnecker, 2003Bonnell, 2017Traficante, 2018Urquhart, 2019Svoboda}). The extreme pressures and temperatures within these cores facilitate the collapse of material, leading to the creation of protostellar objects. The rapid accretion of surrounding material onto these protostars can trigger the release of intense radiation and powerful outflows, establishing an intricate balance between inward gravitational forces and outward pressure. The interplay of physical forces during high-mass star formation contributes to the observed clustering of these stars. These clusters play a crucial role in shaping the subsequent evolution of the stars within them, as well as the galaxies in which they reside (e.g,~\citealt{2003McKee, 2007ZY}).


One of the key components that profoundly influences the high-mass star formation process is the filamentary structure prevalent in molecular clouds. These elongated, thread-like structures, have been observed in various molecular tracers and continuum emission, indicating their essential role in the formation and distribution of high-mass stars assisting in the flow of material onto individual cores (e.g,~\citealt{2008Goldsmith, 2009Myers, 2010Andre, 2010Schneider}). Accretion is a central process in the early stages of star formation. By examining how mass flows onto a core, the mechanisms driving their growth can be investigated. Understanding the interplay between accretion, radiation pressure, and other physical processes provides a clearer picture of how these massive objects form from their natal material. 
Filamentary structures have been found on many spatial scales, a full review of the filamentary ISM can be found in (e.g,~\citealt{2023Hacar, 2020Schisano}). Notable Galactic scale structures, extended up to tens and even more than hundreds of parsecs, include "Maggie" \citep{2022Syed}, "Nessie" \citep{2010Jackson}, and the Radcliffe wave \citep{2020Alves}. On smaller scales, the filamentary structures prevalent in molecular clouds and their surrounding have been studied too, some examples include Serpens South \citep{2013Kirk}, G035.39 00.33 \citep{2014Henshaw} and infrared dark cloud G28.3 \citep{2020Beuther}; with mass accretion estimates on the order of 10$^{-5}$~M$_{\sun}$yr$^{-1}$ for all three studies. Previous examples of flow rate analysis carried out with the Atacama Large Millimetre Array (ALMA), are seen in the works from \cite{2013Peretto}, \cite{2021Sanhueza}, \cite{2022Redaelli}, and \cite{2023O} all giving estimates on the order of 10$^{-4}$~M$_{\sun}$yr$^{-1}$.

Evidence for both radial and longitudinal flows have been observed, each representing different kinds of material transport. Radial traces flows from the environment onto the filament and help build up its mass, however longitudinal flows trace flows along the filament and onto cores, building up the core mass. The kinematic molecular gas study done by \cite{2014Tackenberg} complimenting work done on 16 high-mass star-forming regions from the \textit{Herschel} key project The Earliest Phases of Star formation (EPoS) shows that profiles perpendicular to the filament have almost constant velocities and that the velocity gradient occurs predominantly along the filament. Regions often have unique filamentary structures, but in most cases, velocity gradients can be identified along these filamentary structures towards the central hubs of clumps, which allows mass accretion estimates to be calculated.

Looking at derived parameters throughout the stages of evolution can provide constraints for theoretical models.
It is especially important to investigate all aspects of the high-mass star formation process throughout the complete evolutionary sequence so that we can compare and analyse how these results change through the lifetimes of (proto)stars.

In this paper we use a subset of the regions from the ALMA Evolutionary study of High Mass Protocluster Formation in the Galaxy (ALMAGAL) survey (Molinari et al. in prep) (see Sect. \ref{sect:almagalsurvey}) to investigate properties of flow rates, focusing on longitudinal flows along filamentary structures towards the high-mass cores. Making use of selected spectral lines we estimate flow rates onto individual cores as a function of the evolutionary stage (see Sect. \ref{subsect:evolution}), distance from the cores or core masses.  Qualitative and quantitative results are discussed in the context of high-mass star-forming clusters.

The structure of the paper is as follows: the survey introduction and overview are given in Sect. \,\ref{sect:almagalsurvey}. In Sect. \,\ref{sect:sample} we introduce the sample along with how and why the regions were selected. In Sect. \,\ref{sect:initialanalysis} we investigate the selected cores in more detail looking at signs of potential outflow signatures and line properties for velocity estimation. Details of how the flow rate calculation is done with detailed parameter descriptions are presented in Sect. \,\ref{sect:flow rates}. Sect. \,\ref{sect:results} presents the results of this calculation on our sub-sample before discussions in Sect. \,\ref{sect:discussion} including an expansion to theory and a wider context. We draw our conclusions in Sect.\,\ref{sect:conclusion} and discuss opportunities for future work.

\section{Sample selection}
For this analysis, we choose a smaller subset of 100 regions for a focused study on flow rates and the relationship between them and other core properties. These sources were selected visually based on having strong continuum and line emission so that the initial sample includes 25 regions from each of the 4 evolutionary stages (quiescent, protostellar, Young Stellar Object (YSO), and \HII~Region, see Sect. \ref{subsect:evolution}).
\label{sect:sample}
\subsection{ALMAGAL survey details}
\label{sect:almagalsurvey}
\begin{figure}[]
    \centering
    \includegraphics[width=0.48\textwidth, height=0.34\textheight]{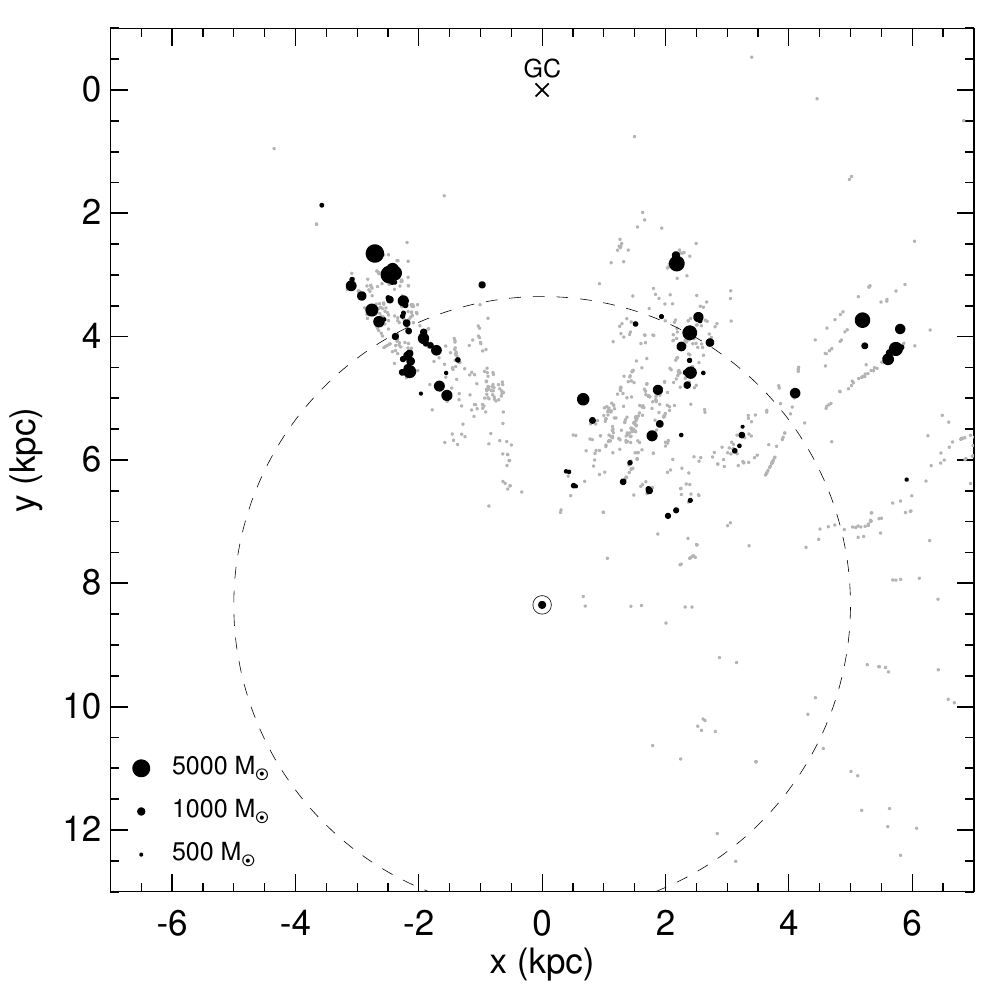}
    \caption{Source distribution for the regions in our ALMAGAL sub-sample is shown as black dots. The size of the markers scales with the masses of the ALMAGAL clumps. Grey dots are the rest of the ALMAGAL survey and the dashed line is a heliocentric distance circle at 5~kpc.}
    \label{fig:survey}
\end{figure}
\begin{table}[]
    \centering
    \caption{Observational parameters}
    \begin{tabular}{lc}\hline\hline
         Parameter & Value \\\hline
         Mean beam size & 0.8\arcsec\\
         Continuum RMS & $\sim$0.15 mJy/beam\\
         H$_2$CO (3$_{0,3}$ - 2$_{0,2}$) RMS & $\sim$9 mJy/beam\\
         SO (6$_5$ - 5$_4$) RMS & $\sim$5 mJy/beam\\
         Velocity resolution & $\sim$~488kHz $\sim$~0.7 kms$^{-1}$\\\hline
         \end{tabular}
    
    \label{tab:technical}
\end{table}

The ALMAGAL survey (2019.1.00195.L; PIs: Sergio Molinari, Peter Schilke, Cara Battersby, Paul Ho) is a large program approved during in ALMA Cycle 7. The ALMAGAL targets consist of 1013 compact dense clumps, covering different evolutionary stages, the majority being selected from the Herschel Hi-GAL survey (\citealt{molinari2010, Elia2017, Elia2021}), with $\sim$ 100 regions come from the Red MSX Source (RMS) survey  (\citealt{2005Hoare, 2007Urquhart, lumsden2013}). The 1017 targets are spread across the Galaxy. Figure \ref{fig:survey} shows their distribution in the face-on view of the Galactic plane.
The 1017 regions were observed in ALMA band 6 at frequencies from 217 to 221~GHz (corresponding to 1.3~mm). Information about the observations, data reduction and image generation are presented and discussed in more detail in the ALMAGAL data reduction paper (Sanchez-Monge et al., in prep). Here, we present a brief description of those aspects relevant for the scientific analysis of this work. The ALMAGAL spectral setup was designed to have four different spectral windows, two of them covering a broad frequency range of 2x 1.875~GHz being sensitive to the continuum emission, as well as many spectral lines at low spectral resolution (1.3~km/s); and two narrower spectral windows (2x 0.468~GHz) aimed at studying specific molecular species (e.g., H$_2$CO, CH$_3$CN) at higher spectral resolution (0.3~km/s). In this work, we make use of the spectral lines H$_2$CO (3$_{0,3}$ - 2$_{0,2}$) at 218.222 GHz and SO SO (6$_5$ - 5$_4$) at 219.949 GHz.
The ALMAGAL observations made use of three different array configurations to observe each source, including two different configurations of the main 12m ALMA array, and observations with the 7m Atacama Compact Array. This allows observations sensitive to angular scales from 0.1\arcsec up to 10\arcsec. The data products used in this work are images with combined data from the 7m array (7M hereafter) and the most compact 12-m array configuration (TM2 hereafter). We note that this work does not make use of the most extended array configurations available within the ALMAGAL project. The resulting images have angular resolutions of 0.5-1.0\arcsec, depending on the distance of the source, which result in similar linear resolutions of 5000~au for all the targets. See Table \ref{tab:technical} for details on typical observational parameters. The entire survey, including full observational details, is described in Molinari et al. (in prep.) while the details of the data-reduction pipeline are in Sanchez-Monge et al. (in prep.).


\begin{figure*}[ht]
\begin{subfigure}{0.5\textwidth}
\includegraphics[width=0.99\textwidth, height=0.3\textheight]{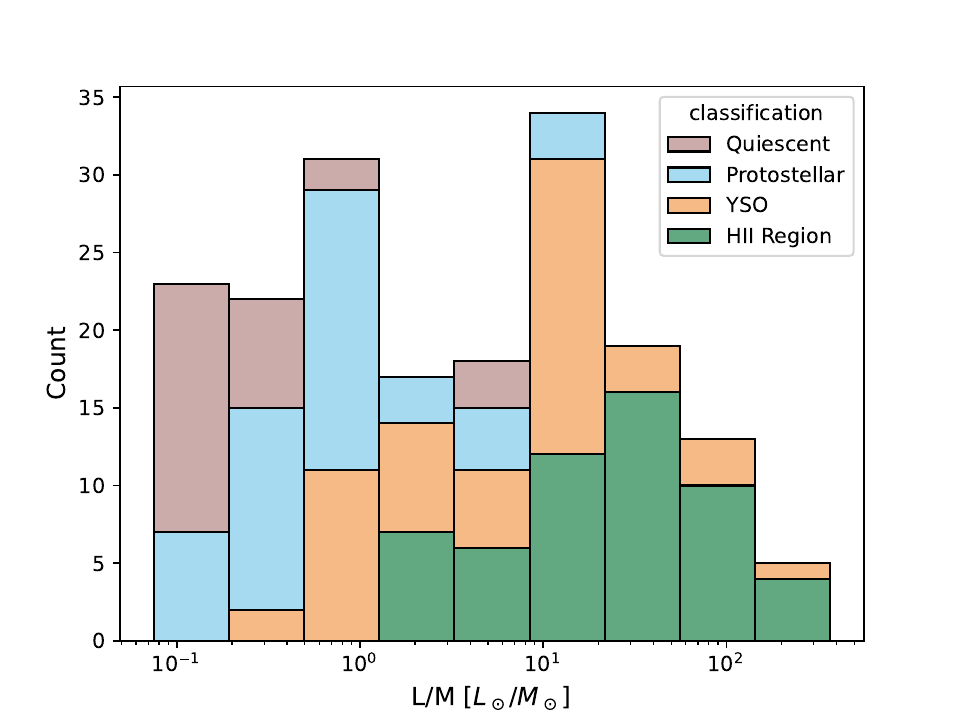} 
\caption{Stacked histogram}
\label{fig:lm_hist}
\end{subfigure}
\begin{subfigure}{0.5\textwidth}
\includegraphics[width=0.99\textwidth, height=0.3\textheight]{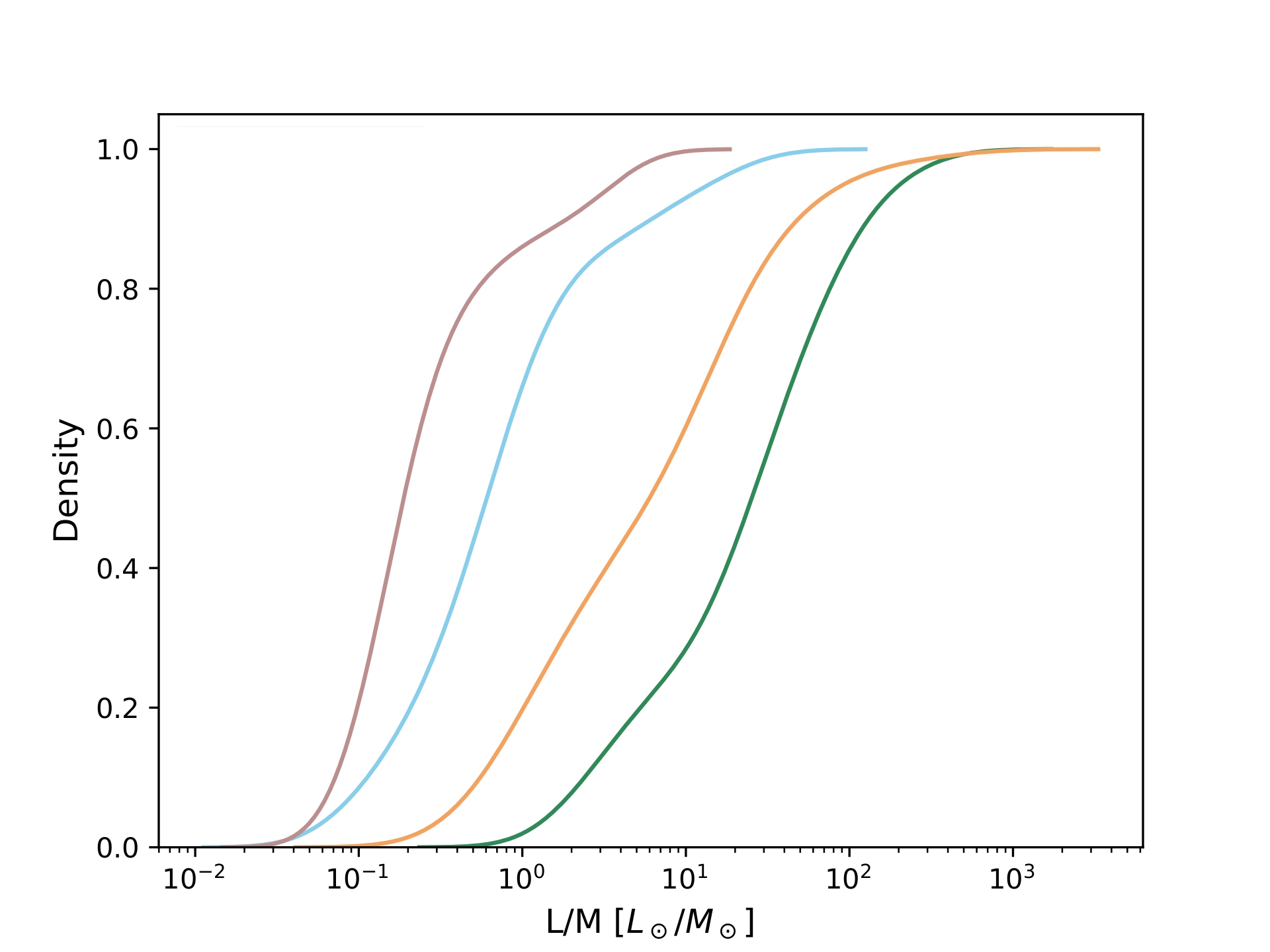}
\caption{Cumulative distribution function}
\label{fig:individual_cdf}
\end{subfigure}
\caption{Colour coding for both figures is on evolutionary stage, legend shown in panel (a). These plots show the sample of 17 quiescent, 23 protostellar, 22 YSO and 25  HII~Regions. (a) Stacked histogram distribution of the luminosity-to-mass ratio for the regions in the ALMAGAL sub-sample being used for this work. (b) Individual cumulative distribution functions (CDF) of regions in each evolutionary stage, generated from kernel density estimates (KDEs) of the data shown in (a).}
\label{fig:evols}
\end{figure*}

\subsection{Sample}
\label{bias}
Looking at our sample compared to the whole survey we check how the mass, luminosity, luminosity-to-mass ratio and distance distributions compared to each other and whether there were signs of any bias. We see no signs of bias in distance, luminosity or luminosity to mass ratio. For the mass we specifically chose regions over 500 M$_{\sun}$ and so this is reflected here. Histograms of these distributions can be found in \ref{fig:surveysample}. For more information on how the survey parameters were calculated we refer to Molinari et. al (in prep.), but as an overview, the distances were derived with the \cite{2021Mege} method using the ALMAGAL spectral cubes and following this the distance dependent quantities were calculated.

\subsection{Evolutionary stage}
\label{subsect:evolution}
Before selecting the regions for the analysis, the ALMAGAL sample is classified by evolutionary sequence. We use the sequence and classification scheme defined by \cite{2022Urquhart}, which divides the sources into four evolutionary stages; Quiescent, Protostellar, Young Stellar Object (YSO), and \HII~Regions. The classification is done by looking at the sources at three wavelengths: Hi-GAL 70~\si{\micro\metre} \citep{Elia2017}, MIPSGAL 24~\si{\micro\metre} \citep{2009Carey}, and GLIMPSE 8~\si{\micro\metre} \citep{2009Churchwell}. Quiescent sources have a central area free of emission at all three wavelengths. For protostellar sources, there is a point source in the 70~\si{\micro\metre} image, potentially a 24~\si{\micro\metre} counterpart but the source is not visible in the 8~\si{\micro\metre} image. A YSO is detected as a point-like source at all 3 wavelengths. \HII~regions also have a point source at all three wavelengths but the source in the 8~\si{\micro\metre} image becomes more extended and `fluffy'. 

Initially, the ALMAGAL sample was cross-matched with the ATLASGAL (\citealt{2009Schuller, 2018Urquhart, 2022Urquhart}) sample to see how many sources overlap and how many classifications can be immediately adopted. Cross-matching on Galactic coordinates with an error margin of 40\arcsec~leads to an initial match of roughly 600 regions out of 1013. The remaining ALMAGAL regions were classified visually according to the same rules as described above.

As an alternative evolutionary indicator, one can also look at the luminosity-to-mass ratio. This ratio increases with time, being very low in the early quiescent stage  (e.g.,~\citealt{2019Molinari, Elia2021}). The distribution of the regions in each classification stage can be seen in Fig. \ref{fig:lm_hist} which shows the peak luminosity-to-mass ratio progression as the cores become more evolved. Figure \ref{fig:lm_hist} enables a comparison of our evolutionary sequence with this evolutionary indicator, and as expected the two classification schemes roughly agree, with some overlap. This can also be seen in Fig. \ref{fig:individual_cdf} showing a cumulative distribution. These classifications are done for the entire cluster-forming clump. However, within individual ALMAGAL regions, \HII~regions, YSO, Protostellar, and even Quiescent regions often coexist. Examples of this include e.g., NGC6334I (e.g.,~\citealt{2005Beuther}), or G29.96+0.02 (e.g.,~\citealt{1998Cesaroni}, ISOSS\,J23053+5953 \citep{2022Gieser}.

\subsection{Core identification}
\begin{figure}[t!]
    \centering
    \includegraphics[width=0.49\textwidth, height=0.34\textheight]{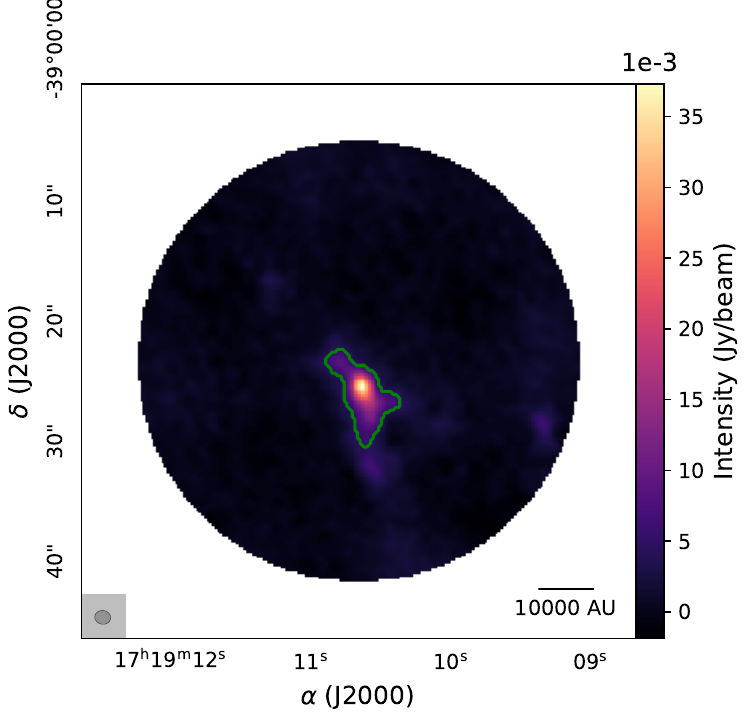}
    \caption{Continuum image of ALMAGAL source AG348.5792-0.9197 from the \texttt{astrodendro} package where the green contours indicate the "leaves" which are the areas we focus on here, the cores.}
    \label{fig:dendro}
\end{figure}
Each source in the subset of our ALMAGAL sample contains up to seven cores identified via the following process. We used the \texttt{astrodendro}\footnote{http://www.dendrograms.org/} program on the continuum data for core identification, obtaining the peak position of these identified cores and estimating their peak and integrated flux density values. The \texttt{astrodendro} package allows us to break down the hierarchical structures in our observational data. The highest hierarchical level for each structure is a "leaf" (i.e a structure with no substructure), these correspond to what we define as a core. We can see an example of a "leaf" in Fig \ref{fig:dendro}, which shows the case of source AG348.5792-0.9197. The three main input parameters of \texttt{astrodendro} are $min\_value$ (the minimum pixel intensity to be considered), $min\_delta$ (the minimum height for any local maximum to be defined as an independent entity), and $min\_npix$ (the minimum number of pixels for a leaf to be defined as an independent entity). We decided to have a large significance level for the cores to be identified so that we were left dealing with just the cores themselves and their structures and not the extended parental cloud. We use $min\_value$ = 5~$\sigma_{cont}$, $min\_delta$ = 5~$\sigma_{cont}$ and $min\_npix$ = beam area. With the combination of our choices of $min\_value$ and $min\_delta$,  all cores have a peak flux density  $\geq$ 10~$\sigma_{cont}$. Here $\sigma_{cont}$ or $\sigma_{line}$ are the rms values of either the continuum image or the spectral cube for the lines being used. Running the analysis with these parameters we identify 203 cores within the 100 regions. Of these 100 regions, five regions had no cores identified with our criteria, so these were removed from the sample leaving an initial 95 regions with 203 cores. The official core catalogue for ALMAGAL calculated on the final data products, including also more extended ALMA configurations, will be available in Coletta et al. (in prep.).

\subsection{Analysis}
\label{sect:initialanalysis}
\begin{figure}[t!]
    \centering
    \includegraphics[width=0.49\textwidth, height=0.34\textheight]{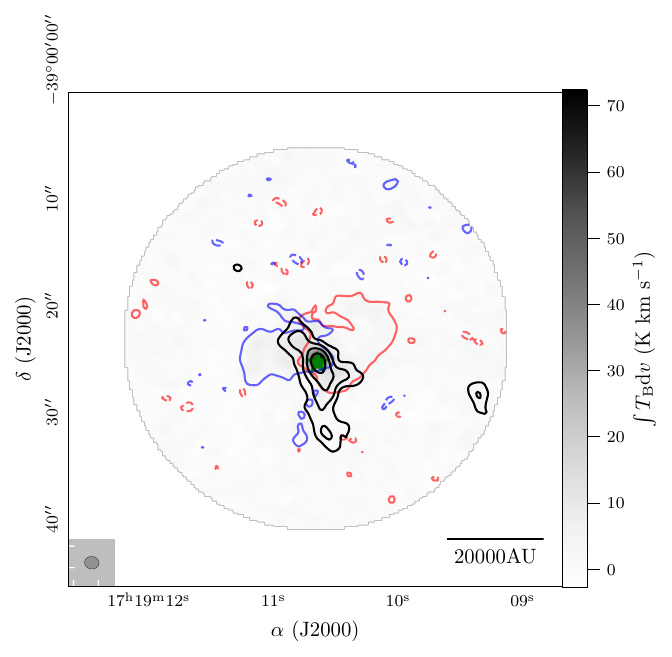}
    \caption{0th moment map of SO (6$_5$ - 5$_4$) in grey-scale for source AG348.5792-0.9197 overlaid with continuum contours in black (levels 3,6,9 $\sigma_{cont}$). A green star to show the peak intensity position of the core. Red and blue contours show the "wings" of the spectral line emission, from 3 to 20~kms$^{-1}$ either side, with respect to the region velocity of rest.}
    \label{fig:outflow}
\end{figure}

We start the analysis with a detailed look into the main lines suitable for identifying outflows, such as SiO (5 - 4) and SO (6$_5$ - 5$_4$) and making Position-Velocity (PV) cuts along the filamentary structures surrounding each core (identified visually from the continuum contours which can be seen in Figs. \ref{fig:outflow} and \ref{fig:momentmap}), in the H$_2$CO (3$_{0,3}$ - 2$_{0,2}$) line.  

\subsubsection{Outflows}
This work focuses primarily on longitudinal flows along filamentary structures. To ensure there was no contamination from any associated outflows we looked at the shock tracers available in the ALMAGAL survey. After a comparison of shock/outflow tracers SiO (5 - 4) and SO (6$_5$ - 5$_4$) it was evident that SO (6$_5$ - 5$_4$) presented the most outflow signatures, manifested as blue- and red-shifted line wing emission ("wing" structures on 0th-moment maps, inspected visually, seen in Fig. \ref{fig:outflow}) (e.g,~\citealt{2016W, 2021VG}). It must be noted that these are not definite detections, just indications of outflows, and this decision was made by visually inspecting the data and results for both lines. We investigated the presence of any red and blue shifted SO (6$_5$ - 5$_4$) emission, before continuing with our analysis. Figure \ref{fig:outflow} shows an example of such an analysis, and we can see signatures of a bipolar outflow from the central core in source AG348.5792-0.9197 (denoted by a green star in the figure). We see red-shifted emission extending to the west and blue-shifted emission to the east, both almost perpendicular to the filamentary structure. With such an analysis, we can focus on the filamentary structures and calculate flow rates along this axis without major outflow contamination.


\subsubsection{Position-Velocity diagrams}

Of all the strong lines available to use in the ALMAGAL spectral setup we decided to cut along the visibly elongated filamentary-like structures in the PV space using H$_2$CO (3$_{0,3}$ - 2$_{0,2}$) due to its intermediate critical density ($\sim$~7$\times$10$^{5}$ cm$^{-3}$, \cite{shirley2015}) that is similar to the densities we expect to trace in our regions. H$_2$CO (3$_{0,3}$ - 2$_{0,2}$) is also a tracer of relatively cool gas ($E_u / k$ ~ 21K) and can be used in combination with another H$_2$CO line as a well-known temperature tracer ($\sim$~100K e.g.,~\citealt{shirley2015, mangum1993, 2007tak, 2021Gieser, 2024I}). Abundance may vary by an order of magnitude over the evolutionary stages \citep{2014Gerner}.

\begin{figure*}[h]
\begin{subfigure}{0.5\textwidth}
\includegraphics[width=0.99\textwidth, height=0.34\textheight]{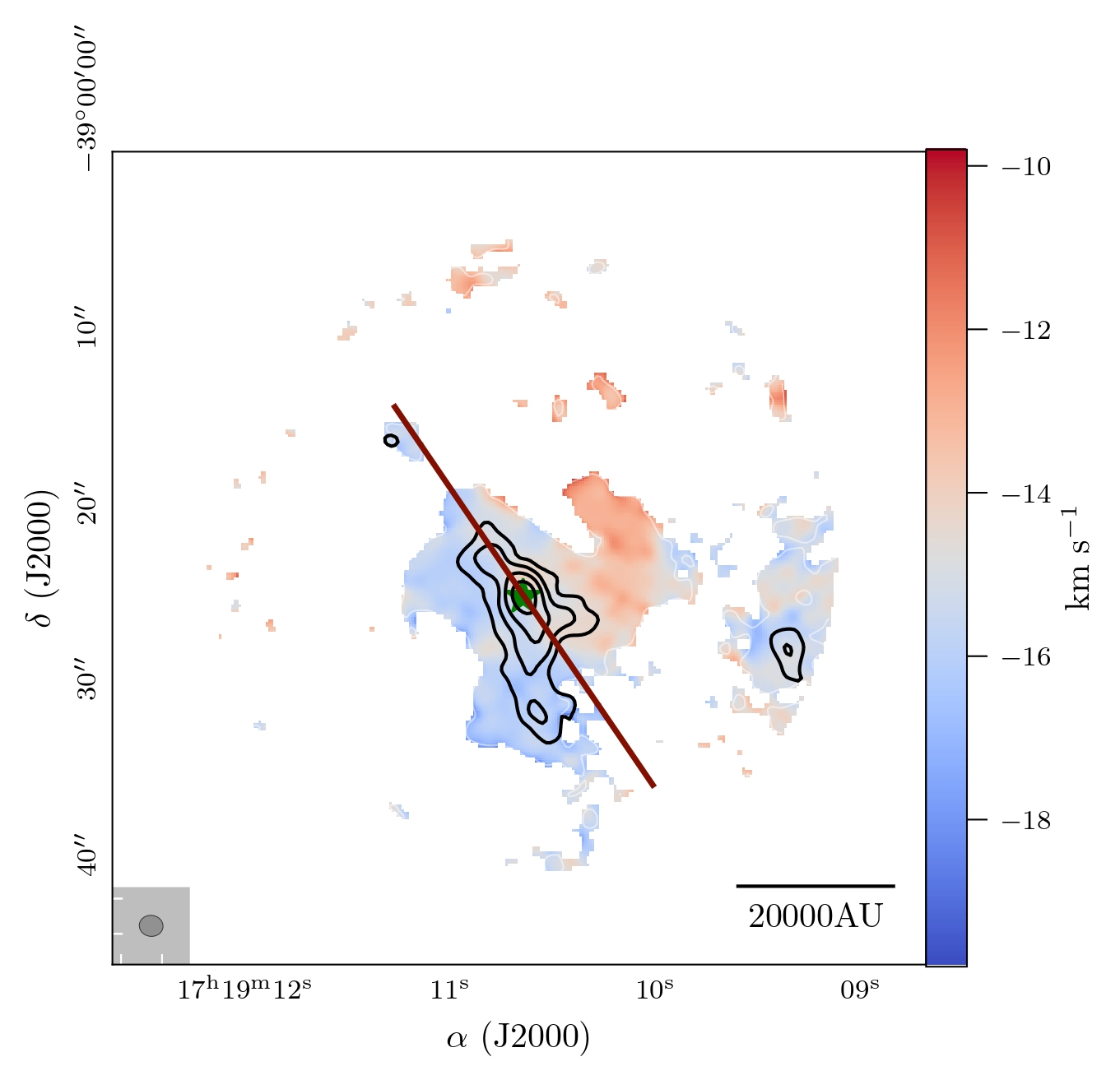} 
\caption{1st moment map.}
\label{fig:momentmap}
\end{subfigure}
\begin{subfigure}{0.5\textwidth}
\includegraphics[width=0.99\textwidth, height=0.3\textheight]{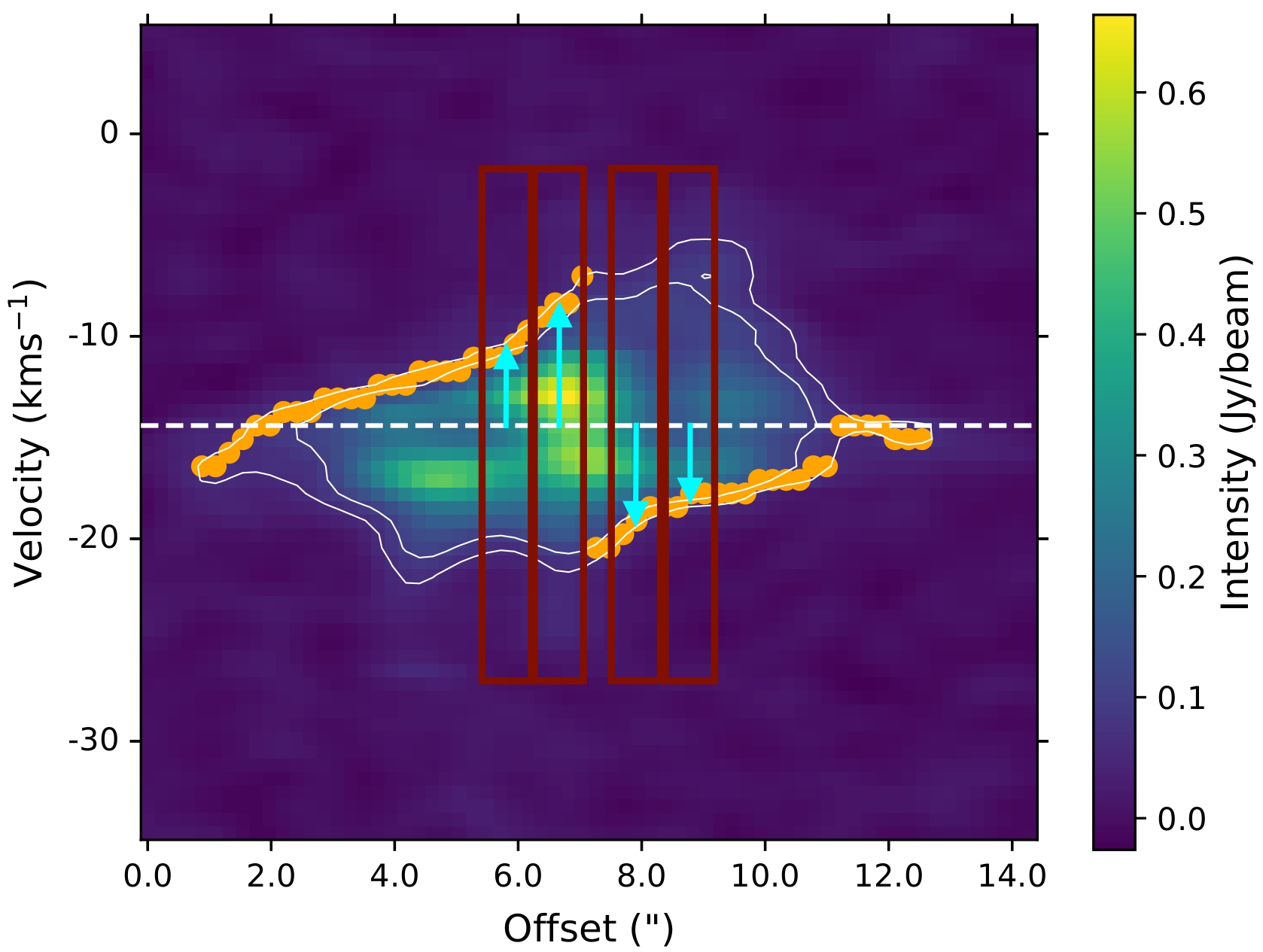}
\caption{Position-Velocity cut.}
\label{fig:extremevelocities}
\end{subfigure}

\caption{\textbf{(a)} 1st moment map of YSO source AG348.5792-0.9197 in H$_2$CO (3$_{0,3}$ - 2$_{0,2}$). The position of the continuum peak flux density is shown by the green star, and continuum contours are shown in black (levels 3,6,9 $\sigma_{cont}$). The red line indicates the axis along which the PV cut was taken. \textbf{(b)} PV cut with 3 and 5 $\sigma_{line}$ contours in white, and the $V_{\mathrm{LSR}}$ of the region shown by the white dashed line. The orange points show the nearest pixels at the 3 $\sigma_{line}$ contours. The red boxes are examples of the areas where we estimate the flow rates across ($\sim$1\arcsec). The peak flux density position of the continuum core is located at the center of each axis.}
\label{fig:pv}
\end{figure*}

The angle at which the PV cut was taken was determined by the outflow signatures and the filamentary structures. To ensure as little contamination from the potential outflows, the cut was made perpendicular to any signatures where possible, whilst keeping the cut inline with the filamentary structure. This is shown in Fig. \ref{fig:pv}. Any cores that did not have suitable emission in the PV cuts were removed from the sample (8 regions, 21 cores).

\section{Flow rates}
\label{sect:flow rates}
After sample selection, core identification, and line analysis the final sample consists of 87 regions with 182 cores in total. Table \ref{final_sample} shows how this sample is split among the evolutionary stages. 

\begin{table}
\caption{Final Sample Distribution}           
\label{final_sample}      
\centering                          
\begin{tabular}{c c c}        
\hline\hline                
Classification & Regions & Cores\\    
\hline                        
Quiescent & 17 & 28\\
Protostellar & 23 & 48\\ 
YSO & 22 & 51\\
\HII~Region & 25 & 55\\
\hline                                   
\end{tabular}
\end{table}

\subsection{Quantifying flow rates}
To estimate the flow rates along the filamentary structures leading toward the cores, we follow the approach outlined in \cite{2020Beuther}.  The mass flows rates $\dot{M}$ are estimated as
\begin{equation}
    \dot M = \Sigma \cdot \Delta v \cdot w,
    \label{flow_equation}
\end{equation}
where $\Sigma$ is the surface density in units of gcm$^{-2}$ (converted from the column density calculated in Sect. \ref{sect:colden}), $\Delta v$ is the velocity difference from the velocity of rest to the 3 $\sigma_{line}$ contour of the PV cut in kms$^{-1}$, considered for the specific flow rate, and $w$ is the width of the area along which the flow rate is measured in AU. The final values of $\dot M$ are converted to M$_{\sun}$yr$^{-1}$. In the following, we describe the parameter determinations in more detail along with details of the calculation (see Appendix \ref{sect:appendixb}).
 
\subsection{Column density}
\label{sect:colden}
Column density maps were made using Equation \ref{colden} (modified black body emission equation from \citealt{2009Schuller}) assuming optically thin dust emission at mm wavelengths \citep{1983H},
\begin{equation}
    N_\mathrm{H_\mathrm{2}} = \frac{F_\mathrm{\nu}R}{B_\mathrm{\nu}(T_\mathrm{D})\Omega \kappa_\mathrm{\nu}\mu m_\mathrm{H}}.
    \label{colden}
\end{equation}
Here $F_\mathrm{\nu}$ is the continuum flux density, $B_\mathrm{\nu}(T_\mathrm{D})$ is the Planck function for a dust temperature $T_\mathrm{D}$ (see Sect. \ref{sect:temp}),  $\Omega$ is the beam solid angle, $\mu$ is the mean molecular weight of the interstellar gas, assumed to be equal to 2.8, and $m_{\mathrm{H}}$ is the mass of a hydrogen atom. We also assumed a gas-to-dust mass ratio $ R = 150 $ (with the inclusion of heavy elements \cite{2011Draine}), and $\kappa_\mathrm{\nu}$ = 0.899~\si{\centi\metre^{2}\gram^{-1}} (interpolated to 1300~\si{\micro\metre} from Table 1, Col. 5 of \cite{ossenkopf1994}). This was used to compute the column density for each pixel from the continuum image. The column density maps will be used to select the specific positions where we compute the mass flow rates. In addition to the column densities, we also estimate the core masses following Equation \ref{mass_estimate} below, 
\begin{equation} 
    M = \frac{d^2 F_\mathrm{\nu}R}{B_\mathrm{\nu}(T_\mathrm{D})\kappa_\mathrm{\nu}},
    \label{mass_estimate}
\end{equation}
(e.g.,~\cite{2009Schuller}). Here we use the same values for $\kappa_\mathrm{\nu}$ and the gas-to-dust ratio, and $d$ is the distance to the source, $F_\mathrm{\nu}$ is taken from the integrated flux density from the cores (leaves) from the \texttt{astrodendro} analysis. Details of how we use these values can be found in Sect. \ref{sect:coremass}.

\subsection{Temperature estimates}
\label{sect:temp}
\begin{figure}[ht!]
    \centering
    \includegraphics[width=0.48\textwidth, height=0.28\textheight]{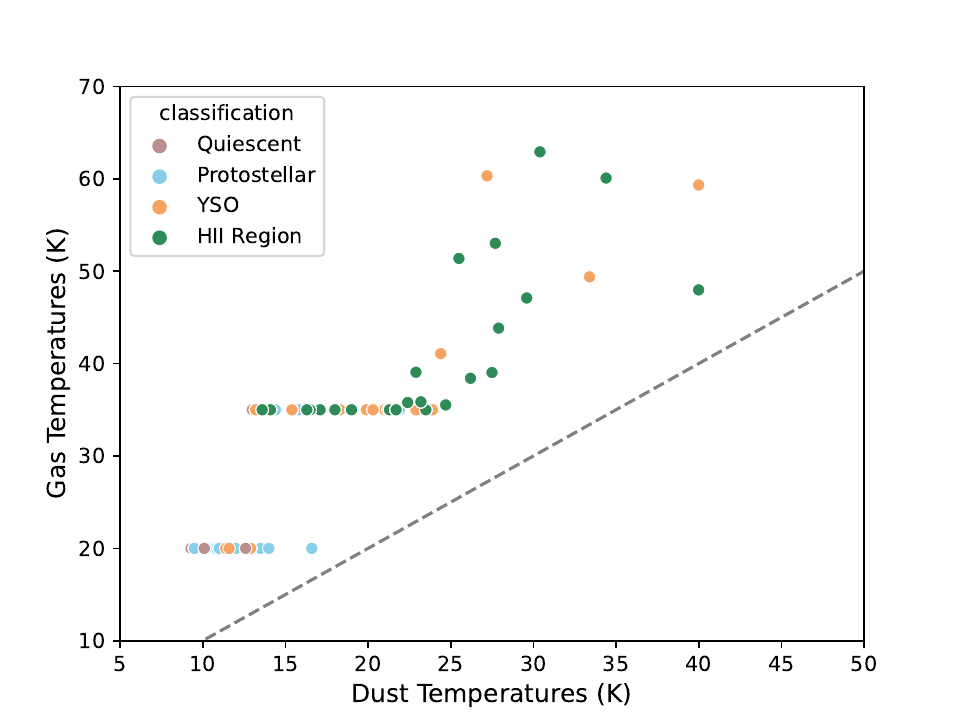}
    \caption{The \textit{Herschel} derived dust temperature and the gas temperature plot against each other colour-coded in evolutionary stage. The grey dashed line shows the temperature equivalence line between the dust and the gas temperatures.}
    \label{fig:temp_comparision}
\end{figure}
Different possibilities exist to estimate the temperatures needed for deriving the column density and mass. Individual estimates per core via molecular line emission of high-density tracers for the ALMAGAL sample will be presented in Jones et al. (in prep.).
While one could use the dust temperatures derived from Hi-GAL \citep{2010Molinari}, they have the disadvantage that they, in general, only sample the colder gas because of the \textit{Herschel} far-infrared wavelength coverage and the large Herschel beam size. In a different approach, \cite{molinari2016} and \cite{traficante2023} calculated temperatures from spectral line emission, where they used luminosity to mass ratio ($ L/M $) values as cut-off points for different temperatures. Following a similar approach, Coletta et al. (in prep.) have estimated temperatures that can be assigned to sources based on their evolutionary stage (indicated by luminosity to mass ratio):
\[ T(L/M)=
\begin{cases}
20~\si{\K} & \mathrm{if}\quad L/M < 1,\\
35~\si{\K} & \mathrm{if}\quad 1\leq L/M < 10,\\
\max(21.1L/M^{0.22}, 35~\si{\K}) & \mathrm{if}\quad L/M > 10.
\end{cases}
\]
Figure \ref{fig:temp_comparision} shows the temperatures derived via these two approaches plotted against each other for each region. The gas temperatures show, as expected from Fig. \ref{fig:individual_cdf}, that quiescent and protostellar sources have temperatures of 20~\si{\K}, while at 35~\si{\K} and above we see mostly YSOs and \HII~Regions, with some exceptions. Comparing this with the dust temperatures where there are some YSO and \HII~Region sources with dust temperature values below 15~\si{\K} leads us to investigate further how this selection would affect the end result. The calculation of the flow rates was therefore done with both the Hi-GAL dust temperatures and the gas temperatures. Comparing the results there were no qualitative and only small quantitative differences ($\sim$ 5-10\%) so we are confident to proceed with the gas temperatures with the reasoning that the dust temperatures are tracing primarily cold gas, whereas the gas temperatures take the warmer protostellar cores into consideration, which aligns more with our aims.

\subsection{Width}
The width parameter $w$, in Equation \ref{flow_equation} is the width of the area along the filament we calculate the flow rate across. We take four areas along each PV cut to allow comparison between results at different offsets from the core. Two inner and two outer positions (0.75\arcsec and 1.75\arcsec~away on either side of the core) excluding the central most 0.5\arcsec~(approximately half a beam size) to avoid contamination as this area is where the flows from all directions are merging. This width is taken as 1\arcsec, the approximate beam size of the data, for all four areas toward each core. An example of this is shown in Fig. \ref{fig:extremevelocities} with the red box showing the width we are looking at to be 1\arcsec. With any conversion to linear distance, we have to take into account the range of distances to the sources in this work. Looking at our range of $\sim$ 2 to 6~kpc (see Fig. \ref{fig:survey}), the more distant sources could have linear width parameters up to a factor three higher. Given that our widths are defined by the (approximate) beam width, at larger distances we capture a larger physical scale, and hence are more likely to have diffuse gas within our source area. This in turn results in more distant sources typically having a lower average column density. While these distance effects may be possible, our derived flow rates show no significant distance dependence, hence the effect should be small.

\subsection{Velocity difference}
To calculate a velocity difference we used the KeplerFit code from \cite{2019Bosco}. The code works by dividing the PV cut into quadrants, taking the strongest (largest intensity) opposing quadrants and reading out the velocity values from the pixels that the specified $\sigma_{line}$ contours go through (here we used 3~$\sigma_{line}$). These points are highlighted in orange in Fig. \ref{fig:extremevelocities}. The velocity measurements are then taken as the difference between the velocity value from the central pixel along the contour confined to each red box in Fig. \ref{fig:extremevelocities}, and the velocity of rest (white dashed line in Fig. \ref{fig:extremevelocities}). In order to use the region's rest velocities as a proxy for the core velocities we compared the rest velocity values to the velocities measured toward the core peak positions in the H$_2$CO 1st moment maps. Comparing the difference between the two, we find that the majority of the values are less than our velocity resolution of 0.6~kms$^{-1}$. In comparison to our median velocity difference measurement $\Delta v$ of 3.4\,km\,s$^{-1}$, this is less than a 20\% error margin. Considering that the ALMA H$_2$CO emission is also affected by missing flux (see Sect. \ref{subsubsect:data} for more information), especially near the peak velocities, this makes the rest velocity a good proxy for the reference velocity.


\subsection{Error analysis}
\label{subsect:errors}

\subsubsection{Data}
\label{subsubsect:data}
Interferometric data without short spacing observations always suffer from missing flux. Regarding the continuum data, comparing to similar studies, e.g., the CORE project in the northern hemisphere with 20 regions was observed with a similar spectral setup and similar baseline ranges, \cite{2018Beuther} estimate 60 to 90\% missing flux across their range of sources. Regarding the spectral line emission, typically the extended emission around the rest velocity is more strongly affected than compact emission offset from the rest velocity. Therefore the lower-level contours (outlined in Figure \ref{fig:extremevelocities}) needed for the PV analysis are not strongly affected. Hence, we are confident that the velocity structure from the H$_2$CO (3$_{0,3}$ - 2$_{0,2}$) line is relatively well recovered. Missing flux has effects on our mass and column density estimates, so we take these values as lower limits.

\subsubsection{Constants}
For the gas-to-dust ratio, we use 150 \citep{2011Draine}. The mean molecular weight of the ISM $\mu$, and the mass of a hydrogen atom $m_{\mathrm{H}}$ both have standard values that were used in our equations \citep{2003Draine}. The dust opacity, $\kappa_\mathrm{\nu}$ was chosen for our conditions and suitable wavelength. For different densities and ice mantels the value could vary up to 30-40\%. Any uncertainties in these parameters here are considered minor compared to the systematic uncertainties discussed above and below, and we are confident that the overall trends we observe will remain consistent.

\subsubsection{Error propagation}
We consider five of our parameters used across this project to have significant uncertainties. These are the flux density, temperature, distance, width, and velocity difference. To calculate the effects this has on our overall results we use Gaussian error propagation for each equation that contains one or more of these parameters. To calculate a mean, standardised error for each flow rate we use mean values combined with the following errors; for the flux density we take 10\% from the calibration uncertainty, for temperature, we take 5~K, for the distance we assume a kinematic distance error of 0.5~kpc, for width we take 0.1\arcsec~for on sky offset discrepancy and finally for the velocity differences we take the spectral resolution of 0.6~km~s$^{-1}$ as the error from the nearest pixel approximation. When combining these we end up with $\pm 50\%$ error margins on our final flow rates. For the core mass, we also estimate roughly $\pm 50\%$ error margins using flux density, temperature, and distance in the Gaussian error propagation.

\subsubsection{Inclination angle} 
We set the inclination angle, $i$, to 0 for the filamentary structures in the plane of the sky. Our input parameters are all affected by the unknown inclination angles. Considering these, Equation \ref{flow_equation} becomes Equation \ref{inclination}, below (full derivation can be found in Appendix A).
\begin{equation}
     \dot M_\mathrm{obs} = \Sigma_{\mathrm{obs}}  \cdot \Delta v_{\mathrm{obs}} \cdot w_{\mathrm{obs}} = \dot M_\mathrm{r} \tan(i)
    \label{inclination}
\end{equation}
Here we are left with a correction factor of $\frac{1}{\tan(i)}$. Hence, the inclination clearly affects the results, meaning our flow rate results have a more narrow distribution in reality as $\tan(i)$ will both increase and decrease with the inclination angle. 
In order to check for the potential spreading of the observed flow rates due to the unknown inclination angle between the filament direction and the observer's line-of-sight, we compute analytically the spreading for an idealised case: We assume a sample of an arbitrary number of filaments with an universal flow rate of $\dot{M} = 10^{-4} \mbox{ M}_\odot \mbox{ yr}^{-1}$ along all filamentary structures of the sample.
We approximate the filaments as cylindrical tube-like structures with a constant and uniform flow rate, hereafter called the ``real'' flow rate $\dot{M}_\mathrm{r}$. The corresponding probability density of the observed flow rates is then given as 
\begin{equation}
\rho_{\dot{M}_\mathrm{obs}}(\dot{M}_\mathrm{obs}) = \frac{1}{\dot{M}_\mathrm{true}} \left(1 + \left(\frac{\dot{M}_\mathrm{obs}}{\dot{M}_\mathrm{r}}\right)^2\right)^{-1.5}.
\end{equation}

\begin{figure}[t!]
   \centering
   \includegraphics[width=0.48\textwidth, height=0.28\textheight]{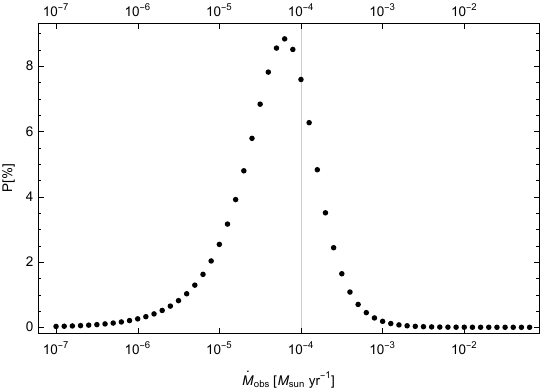} 
   \caption{Theoretical spreading of the observed flow rate due to unknown inclination of the filament for a tube-like cylindrical filament model with a universal flow rate of $\dot{M} = 10^{-4} \mbox{ M}_\odot \mbox{ yr}^{-1}$ (marked as a vertical thin line). The probability distribution is generated for 60 bins with uniform binning width in log space.}
   \label{fig:inclinational-spreading}
\end{figure}

We bin the flow rates in the range from $10^{-6} \mbox{ M}_\odot \mbox{ yr}^{-1}$ to $10^{-2} \mbox{ M}_\odot \mbox{ yr}^{-1}$ into 60 bins with uniform binning width in log space and compute the observational probability by numerically integrating the probability density over the bin.
The final outcome is presented in Fig.~\ref{fig:inclinational-spreading}.

The peak of the probability distribution is quite close to the true flow rate, especially compared to the overall uncertainties of the measurement of the observed flow rates. Also the spread is acceptable with a full width at half maximum of the distribution of quite exactly one order of magnitude in observed flow rate.

In reality, the longer slope toward smaller flow rates will be further reduced (i.e.~will attain a lower probability to be observed) due to the fact that in the simple tube-like model the flow velocity is always aligned with the filament axis and an observer at inclination $i=0$ is assumed to measure zero velocity; in reality, there will be a non-zero velocity in those directions, which in turn reduces the likelihood for observations of the smallest flow rates.
Furthermore, a real sample of filaments will most likely deviate from the assumption of an universal flow rate through all filaments.
This will yield an additional spreading of the distribution of observed flow rates, which is on purpose not taken into account in our analytical model, which analysis focuses on the effect of the unknown inclinations only.

\section{Results}
\label{sect:results}
\begin{figure}[t]
    \centering
    \includegraphics[width=0.48\textwidth, height=0.5\textheight]{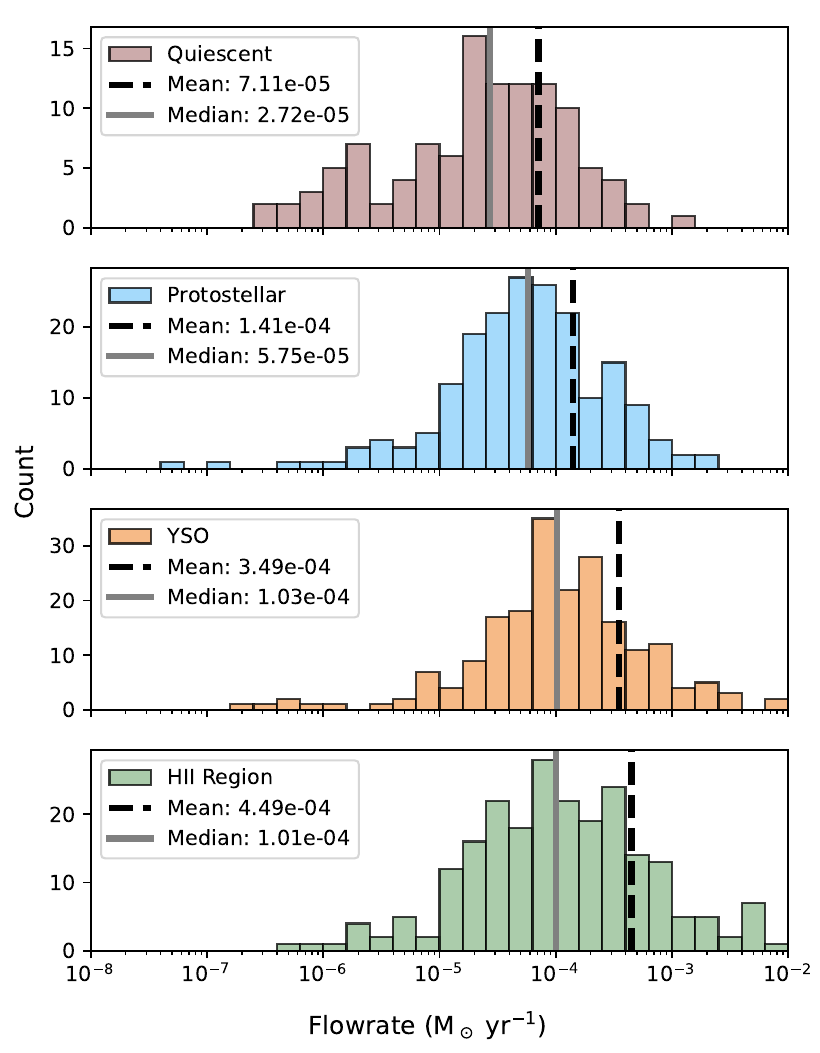}
    \caption{Histograms of the flow rate results for cores in each evolutionary stage quiescent to \HII~Region from top to bottom. The mean and median are shown by the black dashed and grey solid lines, respectively, in each panel.}
    \label{fig:flowhist}
\end{figure}
\begin{figure*}[t]
    \centering
    \includegraphics[width=0.80\textwidth, height=0.34\textheight]{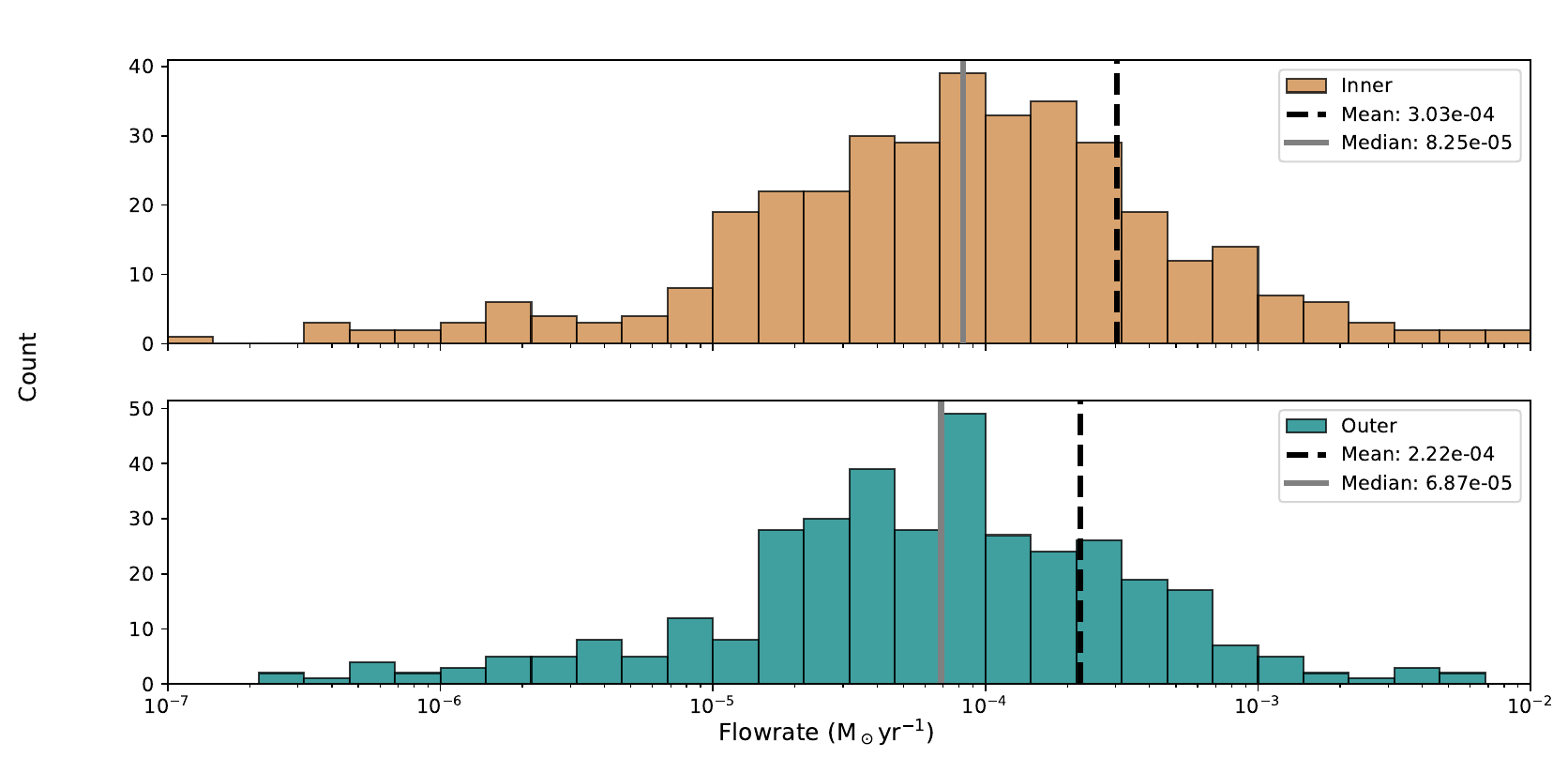}
    \caption{Histograms of the flow rate results in the context of offset from the core. The top panel shows the results from the inner regions and the bottom from those further from the core.}
    \label{fig:innerouter}
\end{figure*}
\begin{figure}[t]
    \centering
    \includegraphics[width=0.49\textwidth, height=0.28\textheight]{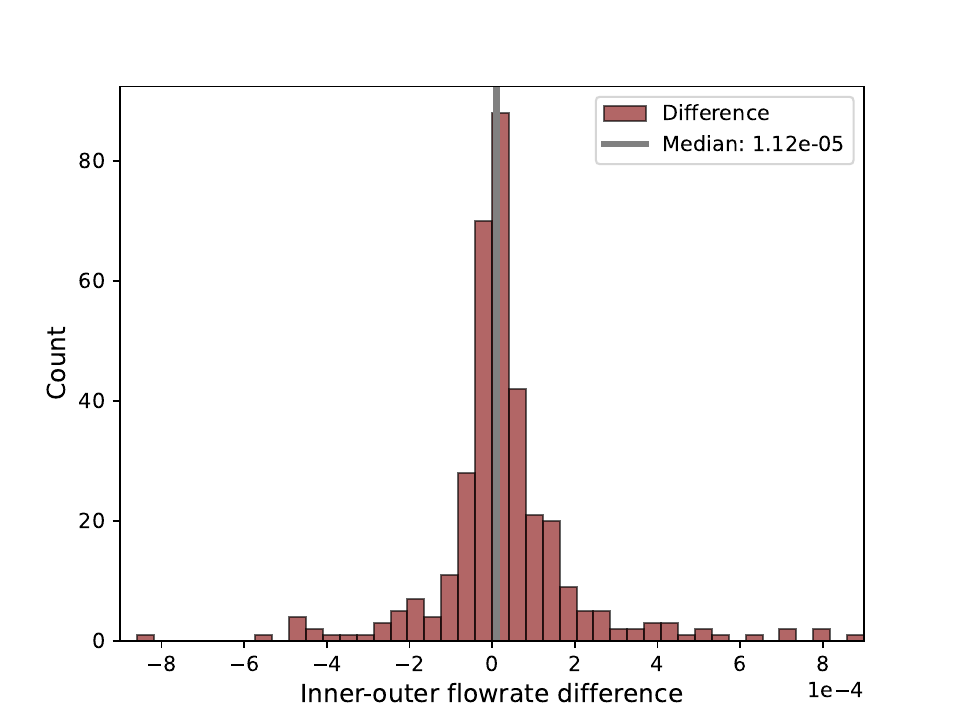}
    \caption{A histogram showing the distribution of the difference between the inner and outer flow rates per core ($\dot{M}_\mathrm{inner}$ - $\dot{M}_\mathrm{outer}$). The median of the distribution is shown as the grey line.}
    \label{fig:difference}
\end{figure}
 
Following the initial analysis and methodology laid out in Sections \ref{sect:initialanalysis} and \ref{sect:flow rates} we present the results for our sample, which using the parameters discussed in Sect. \ref{sect:flow rates} contains 728 measured flow rates. These are mainly constrained between 10$^{-6}~$M$_{\sun}$yr$^{-1}$ and 10$^{-2}~$M$_{\sun}$yr$^{-1}$, with the average values being on the order of 10$^{-4}$~M$_{\sun}$yr$^{-1}$, which is conducive to forming a high-mass star in a few hundred thousand years (e.g., \citealt{2003McKee, 2007Beuther, 2007ZY, 2014T, 2018Motte}). All flow rates can be found in Table \ref{core}. 

Our general assumption for the estimated flow rates is that they are directional toward the cores and hence they are accretion flows. This should certainly be valid for the earlier evolutionary stages: quiescent, protostellar, and YSO. However, that is less clear for the \HII~regions. If one has evolving \HII~regions, those could already be pushing the gas outwards. Hence, the \HII~flow rates are not necessarily accretion flows. An individual classification of each core is beyond the scope of this paper.

\subsection{Statistical testing}
\label{subsect:testing}
To determine the statistical relevance of the results we applied two different, well-known, significance tests, the Kolmogorov-Smirnov (KS) test and the Mann-Whitney U test (\citealt{chakravarti1967handbook,mcknight2010mann}). Both tests are non-parametric, making them suitable for data that may not follow a normal distribution. The KS test focuses on the entire distribution function, while the Mann-Whitney U test looks at the ranks of observations. Both tests generate probability values (p-values), and the interpretation is based on comparing the p-value to a chosen significance level. Here we use 0.05, as used by \cite{chakravarti1967handbook}. The null hypothesis for both tests is the assumption that the samples come from the same distribution (KS) or population (Mann-Whitney U). The KS test generates a p-value, indicating the probability of observing the observed or more extreme differences if the samples come from the same distribution. The Mann-Whitney U test generates a p-value, indicating the probability of observing the calculated U statistic or a more extreme value if the samples come from the same population. 

If our p values from either test are greater than 0.05 then any difference between the distributions is sufficiently small as to be not significant. These results are discussed in sections \ref{sect:evol_results}, \ref{sect:dist} and \ref{sect:coremass} below.

\subsection{Evolutionary stage}
\label{sect:evol_results}
\begin{table}[]
    \centering
    \caption{p-values from the KS and Mann-Whitney U tests.}
    \begin{tabular}{ccc}\hline\hline
        Combination & KS p-value & Mann-Whitney p-value \\\hline
         QP & 0.0457 & 0.7810 \\
         QY & 0.0004 & 0.5725 \\
         QH & 0.0014 & 0.3593 \\
         PY & 0.0406 & 0.8313 \\
         PH & 0.1014 & 0.5501 \\
         YH & 0.6553 & 0.7457 \\\hline
    \end{tabular}
    \label{tab:evol_testing}
    \tablefoot{Combinations are coded with the first letter of the evolutionary stage involved: quiescent, Q; protostellar, P; YSO, Y and \HII~Region, H. Row one denotes the quiescent protostellar combination, and so on.}
\end{table}

Looking at our flow rates in the context of evolutionary stage may tell us about how the accretion process changes as a (proto)star evolves. This result is presented in Figure \ref{fig:flowhist}. The four panels show the distribution for each evolutionary stage from quiescent to the \HII~regions. Considering the $\pm$ 50\% errors, there is a trend between the means and medians of these sub-samples, most notably between Protostellar and YSO sources. In terms of outliers, we have a few that can be seen in Figure \ref{fig:flowhist}. Outliers on the lower end are present in the earlier stages: quiescent and protostellar, and on the higher end in the the more evolved sources. KS and Mann-Whitney tests were done for each combination of the four data sets, the p-value results can be seen in Table \ref{tab:evol_testing}. Using a significance level of 5\%, any p-value above 0.05 tells us there is likely no statistical difference between the data sets. We see that quiescent sources in combination with any of the others are likely not from the same distribution using the KS test however the Mann-Whitney p-value suggests these are not statistically different. By eye, we see there is an increasing trend in the mean or median flow rate through the evolutionary stages. We also note the similarity between these histograms and the distribution in Fig. \ref{fig:inclinational-spreading}, looking at the theoretical spreading due to unknown inclination angle. This gives us an idea that the spread we see in these results is likely partly due to the unknown inclination angle for our observational data.

\subsection{Offset from the core} 
\label{sect:dist}

We now discuss whether the flow rate changes with offset from the core. We look specifically at sections that are 1\arcsec~in width at offsets from the central coordinates of 0.75 and 1.75\arcsec~away on either side of the core, along the filamentary axis. Figure \ref{fig:innerouter} shows the distributions for the flow rates at 0.75\arcsec~(inner) and 1.75\arcsec~(outer) offsets. We see these two distributions have very similar median values but their means are qualitatively different by approximately a factor of two. Again, to determine if these two data sets have a significant difference KS and Mann-Whitney tests were performed (more information in Sect. \ref{subsect:testing}). The p-values from the two tests were 0.0691 and 0.0731 respectively. Using our significance level of 0.05, we cannot reject the null hypothesis that these two distributions are from the same origin. As a further analysis, we looked at the difference between the inner and outer flow rates per core. The distribution can be seen in Fig. \ref{fig:difference} where we can see the distribution centered around 0, with less than 0 meaning the core had higher flow rates further away along the filamentary structure and more than 0 meaning the core has higher flow rates closer to the centre of the core. With a median of 1.12$\times$10$^{-5}~$M$_{\sun}$yr$^{-1}$, we see a trend that the inner flow rates are larger than the outer ones.

\subsection{Core mass}
\label{sect:coremass} 
\begin{figure}[t!]
    \centering
    \includegraphics[width=0.48\textwidth, height=0.28\textheight]{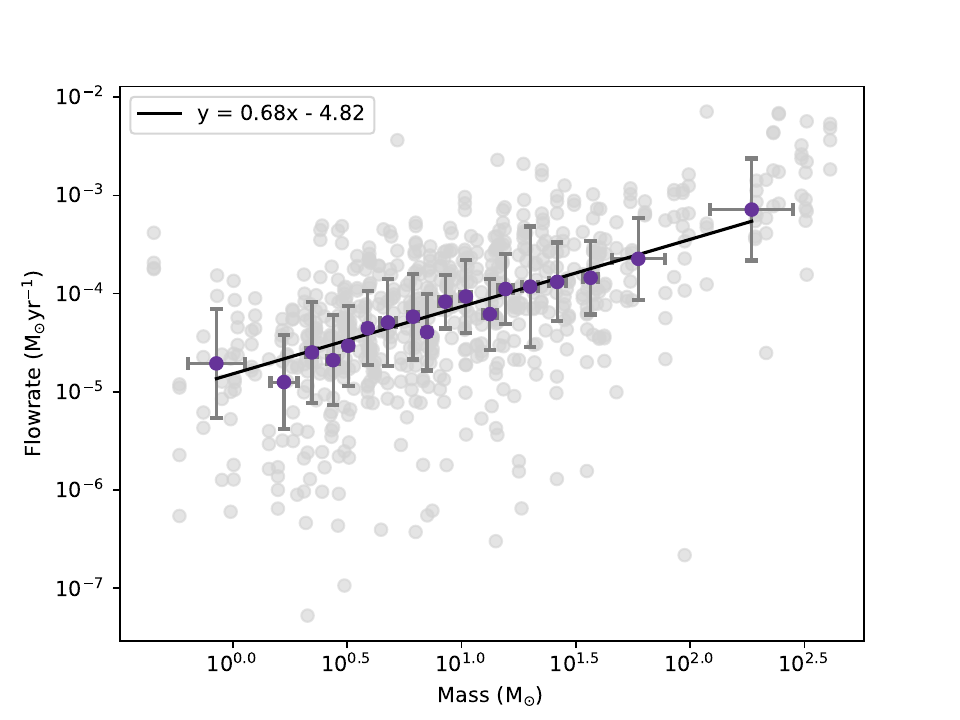}
    \caption{Scatter plot of the results of the whole sample showing flow rate vs. core mass in grey. The purple points are the average flow rate/mass values per bin, with the associated errors. Here, each bin contains the same number of cores. A line of best fit is shown in black.}
    \label{fig:coremass}
\end{figure}

Taking the integrated flux values for each core from the identification analysis (see Sect. \ref{sect:initialanalysis}) we calculated individual core masses, using Equation \ref{mass_estimate}. The distribution of the flow rates vs. core mass is shown in Fig. \ref{fig:coremass} and we see a clear trend between the mass of these cores and the rate at which the material flows onto them. Overlaid in Fig. \ref{fig:coremass} in grey are the same points now binned, first per core (as there are four values per core), and then along the sample. Here we also have a line of best fit through these binned values. This suggests that we have a relationship where the flow rate follows $\sim$M$^{2/3}$. We also looked to see if there was a relationship between these flow rates and the mass of the parental clump and we found no correlation. This indicates that flow rates are largely independent from the parental gas clump and that the found correlation is constrained to the smaller core scales.

As mentioned in Sect. \ref{bias} we looked into whether the sample had any bias' with respect to the whole ALMAGAL sample. Here we want to see if the distance spread, offset or evolutionary stage is causing any unexpected effects. We find no correlation between these flow rates and the distances of these clumps. For the offsets, the distance from the cores, we find that only taking into account the two flow rates closer to the cores gives a steeper relation than the one presented in Fig. \ref{fig:coremass} and if we look at just the flow rates further away from the cores we get a flatter relation. This is not surprising and is in support of Fig. \ref{fig:difference} where we show that the inner flow rates are systematically larger than the outer ones. Comparative higher flow rates closer to the centre and a steeper relation with core mass are supportive that indeed gravity is dominating the infall dynamics. Figures with these relations shown can be found in Appendix \ref{sect:appendixc}.

This is not the first time that a correlation between core mass and accretion rate has been found and or discussed in the literature. \cite{2016Beltran} compiled YSOs with a range of masses and looked at the relationship between their mass and their accretion rates, getting an overall relationship proportional to $\sim$M$^{2}$. During our analysis this relationship was looked at per evolutionary stage to see if in Fig. \ref{fig:coremass} we were seeing any of the evolutionary stages clumped together but this was not the case. We cannot comment on a specific relationship for our YSO values. \cite{2021Clark} discuss what the resulting exponent in this relationship can mean in terms of different mass accretion mechanisms and also the star clusters system mass function. The two accretion mechanisms they discuss are tidal-lobe and Bondi-Hoyle \citep{2001Bonnell}. It is thought that tidal-lobe dominates when the potential of the cluster is still dominated by gas, this mechanism has an exponent of 2/3. The Bondi–Hoyle accretion mechanism dominates when the potential in the cluster is dominated by proto-systems, this mechanism has exponent 2. Our results are in clear agreement with the tidal-lobe accretion mechanism where the potential is dominated by the gas. This is consistent with the ALMAGAL sample covering early evolutionary stages.

\section{Discussion}
\label{sect:discussion}

\subsection{Comparison between low- and high-mass regions}
In this section, we discuss how the flow rates estimated in this work compare to previous studies that quantitatively describe flow rates. The flow rates we present here are comparable to others in the literature for different mass ranges and scales,(e.g,~\citealt{2010Lopez, 2013Duarte, 2013Kirk, 2013Peretto, 2014Henshaw, 2017Traficante, 2020Beuther, 2021Sanhueza, 2022Redaelli}).
Looking at an example from the low mass case, \cite{2013Kirk} uses their Mopra survey of multiple molecular emission lines to look for possible accretion flows onto the central cluster. They present values on the order of 2.8$\times$10$^{-5}~$M$_{\sun}$yr$^{-1}$. For an example of a high mass region, \cite{2014Henshaw} investigates the filamentary structure of an infrared dark cloud G035.39-00.33 in N$_\mathrm{2}$H$^{+}$ and finds mass accretion rates of  7$\times$10$^{-5}$~M$_{\sun}$yr$^{-1}$ with individual filaments feeding individual cores. An example with varying distances from the core is shown in \cite{2020Beuther}, where they look at infrared dark cloud G28.3 using $^{13}$CO and, depending on the distance from the core, they present values around 5$\times$10$^{-5}$~M$_{\sun}$yr$^{-1}$.
If we zoom out and look at larger clumps, \cite{2017Traficante} report mass accretion rates between 0.04$\times$10$^{-3}$ and 2$\times$10$^{-3}$~M$_{\sun}$yr$^{-1}$. They also report seeing an apparent increase in the accretion rate depending on the presence of embedded 24~\si{\micro\metre} sources. This correlates to seeing a difference between our less evolved protostellar sources and more evolved YSO sources.

The results presented here are consistent with the results from the works in the literature mentioned above. The work done by \cite{2013Kirk} in the Serpens South region results are between a factor of 10 and 1000 times smaller than the results we present. If we think about the relationship between flow rate and core mass shown in Sect. \ref{sect:coremass} this is to be expected. Furthermore, comparing low to high-mass star formation, actual accretion rates in the high-mass regime tend to be much higher. There are many competing models discussing the formation timescale of high mass stars (e.g., \citealt{1998Bonnell, 2001Bonnell, 2003McKee, 2007Beuther, 2007Bonnell, 2012Hartmann, 2014T, Motte_2018}) giving approximately 10$^5$ - 10$^6$ years. This then explains that our protostellar and YSO sources, similar to the one studied in \cite{2014Henshaw}, exhibit very similar results. 

\subsection{Comparison to simulations}
Observational studies and theoretical models are extremely complementary to each other for advancing our knowledge in many topics. Here we compare our results to theoretical models that have quantitatively produced accretion flow rates, looking specifically at the work done by \cite{2020Padoan} and \cite{2014Gomez}. 

Looking at the simulations by \cite{2020Padoan}, they produce a sample of roughly 1,500 stars within a volume of 250~pc and study the physical conditions surrounding the sample. The range of this simulation provides a large statistical sample of massive stars, forming realistic distributions of initial conditions. They present mean mass accretion rates on the order of $\sim$10$^{-5}$~M$_{\sun}$yr$^{-1}$ onto the core and they also look at the mass accretion rate 1~pc away from the core and find it increases by an order of magnitude, which agrees with \cite{2017Traficante} on their values for larger scale accretion rates. They also state that their largest values are nearly 10 times higher than these mean values. The range of results we get from our sample agrees with the orders of magnitude discussed in their work. They go on to discuss whether the accretion rate grows systematically with time. We agree with their interpretation that this is not systematic (in their case at the ends of the prestellar phase, in our case throughout our evolutionary sequence).

Turning to \cite{2014Gomez}, they simulate the formation of a molecular cloud from converging gas flows resulting in a dynamic cloud with a lot of substructures, and the cloud grows due to accretion through filamentary structures channelling gas onto the clumps. They look at accretion rates radially along the filament and see a dependence that correlates to changes in the column density profile along the filament. Whilst the method produces filamentary structures the difference in scale makes it hard for a complete interpretation and comparison. Taking into account our work and the examples in this discussion there are definite similarities. The discussion of perpendicular versus parallel flows looking at accretion onto the filament itself and then along towards the central clump is also something discussed by many of these works. In the \cite{2014Gomez} study, they looked at flows both along the filament and perpendicular and even compared themselves to the perpendicular results in \cite{2013Kirk} stating similar values, however also pointing out that their scales are slightly different.

\section{Conclusions}
\label{sect:conclusion}
This work aims to answer the question: what are the properties of accretion flows in high-mass star-forming clusters? This paper presents a subset of the regions from the ALMAGAL survey chosen to investigate the properties of flow rates, focusing specifically on longitudinal flows along filamentary structures towards the central core. A summary of the main results is as follows:
\begin{itemize}
    \item Using calculated column density values and derived velocity differences using the H$_2$CO (3$_{0,3}$ - 2$_{0,2}$) we were able to estimate flow rates for 182 cores from 87 regions of the ALMAGAL survey. We get flow rates on average on the order of $\sim$\flowrate~with error margins of $\pm$50\%. 
    \item We see trends of increasing flow rates through the evolutionary stages, and along the filamentary structure, increasing as we get closer to the central cores. 
    \item  We also see a relationship between the flow rates and the masses of these cores of $\sim$M$^{2/3}$, which supports the tidal-lobe accretion mechanism.
    \item Our results are in line with other observational studies and complementary to theoretical studies in the literature using different methods and mass ranges. Specifically, from the examples discussed, our flow rates are consistent with \cite{Padoan2020}, but we couldn't directly compare to \cite{2014Gomez}.
\end{itemize}
In addition to the conclusions drawn from this project, it is worth noting several supplementary contributions including evolutionary classifications being assigned to the whole ALMAGAL sample to allow for analysis in the context of evolutionary stage, and outflow signatures being detected in this ALMAGAL sub-sample using the SO (6$_5$ - 5$_4$) spectral line; looking at the "wings" of the spectra. Building on the trends we have seen in this work, important next steps would be to see what these relationships look like at both smaller and larger scales, and how these link to each other.

\begin{acknowledgements}
The authors thank Felix Bosco and Daniel Seifried for their discussions during this work and the use of the KeplarFit code by FB. We would also like to thank the referee for the insightful comments and suggestions that improved the paper during the submission process. This research made significant use of \texttt{astrodendro}, a Python package to compute dendrograms of Astronomical data\footnote{http://www.dendrograms.org/},  Astropy:\footnote{http://www.astropy.org} a community-developed core Python package and an ecosystem of tools and resources for astronomy \citep{2013astropy, 2018astropy, 2022astropy}, NumPy \citep{numpy}, matplotlib \citep{matplotlib} and Spectral-Cube \citep{spectralcube}.  This paper makes use of the following ALMA data: ADS/JAO.ALMA2019.1.00195.L. ALMA is a partnership of ESO (representing its member states), NSF (USA), and NINS (Japan), together with NRC (Canada), MOST and ASIAA (Taiwan), and KASI (Republic of Korea), in cooperation with the Republic of Chile. The Joint ALMA Observatory is operated by ESO, AUI/NRAO and NAOJ. RSK acknowledges financial support from the European Research Council via the ERC Synergy Grant ``ECOGAL'' (project ID 855130),  from the German Excellence Strategy via the Heidelberg Cluster of Excellence (EXC 2181 - 390900948) ``STRUCTURES'', and from the German Ministry for Economic Affairs and Climate Action in project ``MAINN'' (funding ID 50OO2206). RSK also thanks for computing resources provided by the Ministry of Science, Research and the Arts (MWK) of the State of Baden-W\"{u}rttemberg through bwHPC and the German Science Foundation (DFG) through grants INST 35/1134-1 FUGG and 35/1597-1 FUGG, and also for data storage at SDS@hd funded through grants INST 35/1314-1 FUGG and INST 35/1503-1 FUGG. Part of this research was carried out at the Jet Propulsion Laboratory, California Institute of Technology, under a contract with the National Aeronautics and Space Administration (80NM0018D0004). RK acknowledges financial support via the Heisenberg Research Grant funded by the Deutsche Forschungsgemeinschaft (DFG, German Research Foundation) under grant no.~KU 2849/9, project no.~445783058. T.L. acknowledges support from the National Key R\&D Program of China (No. 2022YFA1603100); the National Natural Science Foundation of China (NSFC), through grants No. 12073061 and No. 12122307; the international partnership program of the Chinese Academy of Sciences, through grant No. 114231KYSB20200009; and the Shanghai Pujiang Program 20PJ1415500. SW gratefully acknowledges funding via the Collaborative Research Center 1601 (sub-project A5) funded by the German Science Foundation (DFG). A.S.-M.\ acknowledges support from the RyC2021-032892-I grant funded by MCIN/AEI/10.13039/501100011033 and by the European Union `Next GenerationEU’/PRTR, as well as the program Unidad de Excelencia María de Maeztu CEX2020-001058-M, and support from the PID2020-117710GB-I00 (MCI-AEI-FEDER, UE). PS was partially supported by a Grant-in-Aid for Scientific Research (KAKENHI Number JP22H01271 and JP23H01221) of JSPS. GAF  gratefully acknowledges funding via the Collaborative Research Center 1601 (sub-project A2) funded by the German Science Foundation (DFG) and from the University of Cologne through its Global Faculty Program.
\end{acknowledgements}

\bibliographystyle{aa} 
\bibliography{wells_2024.bib}

\begin{thebibliography}{94}
\expandafter\ifx\csname natexlab\endcsname\relax\def\natexlab#1{#1}\fi

\bibitem[{{Alves} {et~al.}(2020){Alves}, {Zucker}, \& {Goodman}}]{2020Alves}
{Alves}, J., {Zucker}, C., \& {Goodman}, A. 2020, Nature, 578

\bibitem[{{Andr{\'e}} {et~al.}(2010){Andr{\'e}}, {Men'shchikov}, {Bontemps},
  {K{\"o}nyves}, {Motte}, {Schneider}, {Didelon}, {Minier}, {Saraceno},
  {Ward-Thompson}, {di Francesco}, {White}, {Molinari}, {Testi}, {Abergel},
  {Griffin}, {Henning}, {Royer}, {Mer{\'\i}n}, {Vavrek}, {Attard},
  {Arzoumanian}, {Wilson}, {Ade}, {Aussel}, {Baluteau}, {Benedettini},
  {Bernard}, {Blommaert}, {Cambr{\'e}sy}, {Cox}, {di Giorgio}, {Hargrave},
  {Hennemann}, {Huang}, {Kirk}, {Krause}, {Launhardt}, {Leeks}, {Le Pennec},
  {Li}, {Martin}, {Maury}, {Olofsson}, {Omont}, {Peretto}, {Pezzuto}, {Prusti},
  {Roussel}, {Russeil}, {Sauvage}, {Sibthorpe}, {Sicilia-Aguilar}, {Spinoglio},
  {Waelkens}, {Woodcraft}, \& {Zavagno}}]{2010Andre}
{Andr{\'e}}, P., {Men'shchikov}, A., {Bontemps}, S., {et~al.} 2010, \aap, 518,
  L102

\bibitem[{{Arce} {et~al.}(2007){Arce}, {Shepherd}, {Gueth}, {Lee}, {Bachiller},
  {Rosen}, \& {Beuther}}]{2007Arce}
{Arce}, H.~G., {Shepherd}, D., {Gueth}, F., {et~al.} 2007, in Protostars and
  Planets V, ed. B.~{Reipurth}, D.~{Jewitt}, \& K.~{Keil}, 245

\bibitem[{{Astropy Collaboration} {et~al.}(2022){Astropy Collaboration},
  {Price-Whelan}, {Lim}, {Earl}, {Starkman}, {Bradley}, {Shupe}, {Patil},
  {Corrales}, {Brasseur}, {N{\"o}the}, {Donath}, {Tollerud}, {Morris},
  {Ginsburg}, {Vaher}, {Weaver}, {Tocknell}, {Jamieson}, {van Kerkwijk},
  {Robitaille}, {Merry}, {Bachetti}, {G{\"u}nther}, {Aldcroft},
  {Alvarado-Montes}, {Archibald}, {B{\'o}di}, {Bapat}, {Barentsen},
  {Baz{\'a}n}, {Biswas}, {Boquien}, {Burke}, {Cara}, {Cara}, {Conroy},
  {Conseil}, {Craig}, {Cross}, {Cruz}, {D'Eugenio}, {Dencheva}, {Devillepoix},
  {Dietrich}, {Eigenbrot}, {Erben}, {Ferreira}, {Foreman-Mackey}, {Fox},
  {Freij}, {Garg}, {Geda}, {Glattly}, {Gondhalekar}, {Gordon}, {Grant},
  {Greenfield}, {Groener}, {Guest}, {Gurovich}, {Handberg}, {Hart},
  {Hatfield-Dodds}, {Homeier}, {Hosseinzadeh}, {Jenness}, {Jones}, {Joseph},
  {Kalmbach}, {Karamehmetoglu}, {Ka{\l}uszy{\'n}ski}, {Kelley}, {Kern},
  {Kerzendorf}, {Koch}, {Kulumani}, {Lee}, {Ly}, {Ma}, {MacBride}, {Maljaars},
  {Muna}, {Murphy}, {Norman}, {O'Steen}, {Oman}, {Pacifici}, {Pascual},
  {Pascual-Granado}, {Patil}, {Perren}, {Pickering}, {Rastogi}, {Roulston},
  {Ryan}, {Rykoff}, {Sabater}, {Sakurikar}, {Salgado}, {Sanghi}, {Saunders},
  {Savchenko}, {Schwardt}, {Seifert-Eckert}, {Shih}, {Jain}, {Shukla}, {Sick},
  {Simpson}, {Singanamalla}, {Singer}, {Singhal}, {Sinha}, {Sip{\H{o}}cz},
  {Spitler}, {Stansby}, {Streicher}, {{\v{S}}umak}, {Swinbank}, {Taranu},
  {Tewary}, {Tremblay}, {de Val-Borro}, {Van Kooten}, {Vasovi{\'c}}, {Verma},
  {de Miranda Cardoso}, {Williams}, {Wilson}, {Winkel}, {Wood-Vasey}, {Xue},
  {Yoachim}, {Zhang}, {Zonca}, \& {Astropy Project Contributors}}]{2022astropy}
{Astropy Collaboration}, {Price-Whelan}, A.~M., {Lim}, P.~L., {et~al.} 2022,
  \apj, 935, 167

\bibitem[{{Astropy Collaboration} {et~al.}(2018){Astropy Collaboration},
  {Price-Whelan}, {Sip{\H{o}}cz}, {G{\"u}nther}, {Lim}, {Crawford}, {Conseil},
  {Shupe}, {Craig}, {Dencheva}, {Ginsburg}, {VanderPlas}, {Bradley},
  {P{\'e}rez-Su{\'a}rez}, {de Val-Borro}, {Aldcroft}, {Cruz}, {Robitaille},
  {Tollerud}, {Ardelean}, {Babej}, {Bach}, {Bachetti}, {Bakanov}, {Bamford},
  {Barentsen}, {Barmby}, {Baumbach}, {Berry}, {Biscani}, {Boquien}, {Bostroem},
  {Bouma}, {Brammer}, {Bray}, {Breytenbach}, {Buddelmeijer}, {Burke},
  {Calderone}, {Cano Rodr{\'\i}guez}, {Cara}, {Cardoso}, {Cheedella}, {Copin},
  {Corrales}, {Crichton}, {D'Avella}, {Deil}, {Depagne}, {Dietrich}, {Donath},
  {Droettboom}, {Earl}, {Erben}, {Fabbro}, {Ferreira}, {Finethy}, {Fox},
  {Garrison}, {Gibbons}, {Goldstein}, {Gommers}, {Greco}, {Greenfield},
  {Groener}, {Grollier}, {Hagen}, {Hirst}, {Homeier}, {Horton}, {Hosseinzadeh},
  {Hu}, {Hunkeler}, {Ivezi{\'c}}, {Jain}, {Jenness}, {Kanarek}, {Kendrew},
  {Kern}, {Kerzendorf}, {Khvalko}, {King}, {Kirkby}, {Kulkarni}, {Kumar},
  {Lee}, {Lenz}, {Littlefair}, {Ma}, {Macleod}, {Mastropietro}, {McCully},
  {Montagnac}, {Morris}, {Mueller}, {Mumford}, {Muna}, {Murphy}, {Nelson},
  {Nguyen}, {Ninan}, {N{\"o}the}, {Ogaz}, {Oh}, {Parejko}, {Parley}, {Pascual},
  {Patil}, {Patil}, {Plunkett}, {Prochaska}, {Rastogi}, {Reddy Janga},
  {Sabater}, {Sakurikar}, {Seifert}, {Sherbert}, {Sherwood-Taylor}, {Shih},
  {Sick}, {Silbiger}, {Singanamalla}, {Singer}, {Sladen}, {Sooley},
  {Sornarajah}, {Streicher}, {Teuben}, {Thomas}, {Tremblay}, {Turner},
  {Terr{\'o}n}, {van Kerkwijk}, {de la Vega}, {Watkins}, {Weaver}, {Whitmore},
  {Woillez}, {Zabalza}, \& {Astropy Contributors}}]{2018astropy}
{Astropy Collaboration}, {Price-Whelan}, A.~M., {Sip{\H{o}}cz}, B.~M., {et~al.}
  2018, \aj, 156, 123

\bibitem[{{Astropy Collaboration} {et~al.}(2013){Astropy Collaboration},
  {Robitaille}, {Tollerud}, {Greenfield}, {Droettboom}, {Bray}, {Aldcroft},
  {Davis}, {Ginsburg}, {Price-Whelan}, {Kerzendorf}, {Conley}, {Crighton},
  {Barbary}, {Muna}, {Ferguson}, {Grollier}, {Parikh}, {Nair}, {Unther},
  {Deil}, {Woillez}, {Conseil}, {Kramer}, {Turner}, {Singer}, {Fox}, {Weaver},
  {Zabalza}, {Edwards}, {Azalee Bostroem}, {Burke}, {Casey}, {Crawford},
  {Dencheva}, {Ely}, {Jenness}, {Labrie}, {Lim}, {Pierfederici}, {Pontzen},
  {Ptak}, {Refsdal}, {Servillat}, \& {Streicher}}]{2013astropy}
{Astropy Collaboration}, {Robitaille}, T.~P., {Tollerud}, E.~J., {et~al.} 2013,
  \aap, 558, A33

\bibitem[{{Beltr{\'a}n} \& {de Wit}(2016)}]{2016Beltran}
{Beltr{\'a}n}, M.~T. \& {de Wit}, W.~J. 2016, \aapr, 24, 6

\bibitem[{{Beuther} {et~al.}(2018){Beuther}, {Mottram}, {Ahmadi}, {Bosco},
  {Linz}, {Henning}, {Klaassen}, {Winters}, {Maud}, {Kuiper}, {Semenov},
  {Gieser}, {Peters}, {Urquhart}, {Pudritz}, {Ragan}, {Feng}, {Keto},
  {Leurini}, {Cesaroni}, {Beltran}, {Palau}, {S{\'a}nchez-Monge},
  {Galvan-Madrid}, {Zhang}, {Schilke}, {Wyrowski}, {Johnston}, {Longmore},
  {Lumsden}, {Hoare}, {Menten}, \& {Csengeri}}]{2018Beuther}
{Beuther}, H., {Mottram}, J.~C., {Ahmadi}, A., {et~al.} 2018, \aap, 617, A100

\bibitem[{{Beuther} {et~al.}(2005){Beuther}, {Thorwirth}, {Zhang}, {Hunter},
  {Megeath}, {Walsh}, \& {Menten}}]{2005Beuther}
{Beuther}, H., {Thorwirth}, S., {Zhang}, Q., {et~al.} 2005, \apj, 627, 834

\bibitem[{{Beuther} {et~al.}(2020){Beuther}, {Wang}, {Soler}, {Linz},
  {Henshaw}, {Vazquez-Semadeni}, {Gomez}, {Ragan}, {Henning}, {Glover}, {Lee},
  \& {G{\"u}sten}}]{2020Beuther}
{Beuther}, H., {Wang}, Y., {Soler}, J., {et~al.} 2020, \aap, 638, A44

\bibitem[{{Beuther} {et~al.}(2007){Beuther}, {Zhang}, {Bergin}, {Sridharan},
  {Hunter}, \& {Leurini}}]{2007Beuther}
{Beuther}, H., {Zhang}, Q., {Bergin}, E.~A., {et~al.} 2007, \aap, 468, 1045

\bibitem[{{Bonnell} {et~al.}(2003){Bonnell}, {Bate}, \& {Vine}}]{2003Bonnell}
{Bonnell}, I.~A., {Bate}, M.~R., \& {Vine}, S.~G. 2003, \mnras, 343, 413

\bibitem[{{Bonnell} {et~al.}(1998){Bonnell}, {Bate}, \&
  {Zinnecker}}]{1998Bonnell}
{Bonnell}, I.~A., {Bate}, M.~R., \& {Zinnecker}, H. 1998, \mnras, 298, 93

\bibitem[{{Bonnell} {et~al.}(2001){Bonnell}, {Clarke}, {Bate}, \&
  {Pringle}}]{2001Bonnell}
{Bonnell}, I.~A., {Clarke}, C.~J., {Bate}, M.~R., \& {Pringle}, J.~E. 2001,
  \mnras, 324, 573

\bibitem[{{Bonnell} {et~al.}(2007){Bonnell}, {Larson}, \&
  {Zinnecker}}]{2007Bonnell}
{Bonnell}, I.~A., {Larson}, R.~B., \& {Zinnecker}, H. 2007, in Protostars and
  Planets V, ed. B.~{Reipurth}, D.~{Jewitt}, \& K.~{Keil}, 149

\bibitem[{{Bosco} {et~al.}(2019){Bosco}, {Beuther}, {Ahmadi}, {Mottram},
  {Kuiper}, {Linz}, {Maud}, {Winters}, {Henning}, {Feng}, {Peters}, {Semenov},
  {Klaassen}, {Schilke}, {Urquhart}, {Beltr{\'a}n}, {Lumsden}, {Leurini},
  {Moscadelli}, {Cesaroni}, {S{\'a}nchez-Monge}, {Palau}, {Pudritz},
  {Wyrowski}, \& {Longmore}}]{2019Bosco}
{Bosco}, F., {Beuther}, H., {Ahmadi}, A., {et~al.} 2019, \aap, 629, A10

\bibitem[{{Bressert} {et~al.}(2010){Bressert}, {Bastian}, {Gutermuth},
  {Megeath}, {Allen}, {Evans}, {Rebull}, {Hatchell}, {Johnstone}, {Bourke},
  {Cieza}, {Harvey}, {Merin}, {Ray}, \& {Tothill}}]{2010Bressert}
{Bressert}, E., {Bastian}, N., {Gutermuth}, R., {et~al.} 2010, \mnras, 409, L54

\bibitem[{{Carey} {et~al.}(2009){Carey}, {Noriega-Crespo}, {Mizuno}, {Shenoy},
  {Paladini}, {Kraemer}, {Price}, {Flagey}, {Ryan}, {Ingalls}, {Kuchar},
  {Pinheiro Gon{\c{c}}alves}, {Indebetouw}, {Billot}, {Marleau}, {Padgett},
  {Rebull}, {Bressert}, {Ali}, {Molinari}, {Martin}, {Berriman}, {Boulanger},
  {Latter}, {Miville-Deschenes}, {Shipman}, \& {Testi}}]{2009Carey}
{Carey}, S.~J., {Noriega-Crespo}, A., {Mizuno}, D.~R., {et~al.} 2009, \pasp,
  121, 76

\bibitem[{{Cesaroni} {et~al.}(1998){Cesaroni}, {Hofner}, {Walmsley}, \&
  {Churchwell}}]{1998Cesaroni}
{Cesaroni}, R., {Hofner}, P., {Walmsley}, C.~M., \& {Churchwell}, E. 1998,
  \aap, 331, 709

\bibitem[{Chakravarti {et~al.}(1967)Chakravarti, Laha, \&
  Roy}]{chakravarti1967handbook}
Chakravarti, I.~M., Laha, R.~G., \& Roy, J. 1967, (No Title)

\bibitem[{{Churchwell} {et~al.}(2009){Churchwell}, {Babler}, {Meade},
  {Whitney}, {Benjamin}, {Indebetouw}, {Cyganowski}, {Robitaille}, {Povich},
  {Watson}, \& {Bracker}}]{2009Churchwell}
{Churchwell}, E., {Babler}, B.~L., {Meade}, M.~R., {et~al.} 2009, \pasp, 121,
  213

\bibitem[{{Clark} \& {Whitworth}(2021)}]{2021Clark}
{Clark}, P.~C. \& {Whitworth}, A.~P. 2021, \mnras, 500, 1697

\bibitem[{{Draine}(2003)}]{2003Draine}
{Draine}, B.~T. 2003, \apj, 598, 1017

\bibitem[{{Draine}(2011)}]{2011Draine}
{Draine}, B.~T. 2011, {Physics of the Interstellar and Intergalactic Medium}

\bibitem[{{Duarte-Cabral} {et~al.}(2013){Duarte-Cabral}, {Bontemps}, {Motte},
  {Hennemann}, {Schneider}, \& {Andr{\'e}}}]{2013Duarte}
{Duarte-Cabral}, A., {Bontemps}, S., {Motte}, F., {et~al.} 2013, \aap, 558,
  A125

\bibitem[{{Elia} {et~al.}(2021){Elia}, {Merello}, {Molinari}, {Schisano},
  {Zavagno}, {Russeil}, {M{\`e}ge}, {Martin}, {Olmi}, {Pestalozzi}, {Plume},
  {Ragan}, {Benedettini}, {Eden}, {Moore}, {Noriega-Crespo}, {Paladini},
  {Palmeirim}, {Pezzuto}, {Pilbratt}, {Rygl}, {Schilke}, {Strafella}, {Tan},
  {Traficante}, {Baldeschi}, {Bally}, {di Giorgio}, {Fiorellino}, {Liu},
  {Piazzo}, \& {Polychroni}}]{Elia2021}
{Elia}, D., {Merello}, M., {Molinari}, S., {et~al.} 2021, \mnras, 504, 2742

\bibitem[{{Elia} {et~al.}(2017){Elia}, {Molinari}, {Schisano}, {Pestalozzi},
  {Pezzuto}, {Merello}, {Noriega-Crespo}, {Moore}, {Russeil}, {Mottram},
  {Paladini}, {Strafella}, {Benedettini}, {Bernard}, {Di Giorgio}, {Eden},
  {Fukui}, {Plume}, {Bally}, {Martin}, {Ragan}, {Jaffa}, {Motte}, {Olmi},
  {Schneider}, {Testi}, {Wyrowski}, {Zavagno}, {Calzoletti}, {Faustini},
  {Natoli}, {Palmeirim}, {Piacentini}, {Piazzo}, {Pilbratt}, {Polychroni},
  {Baldeschi}, {Beltr{\'a}n}, {Billot}, {Cambr{\'e}sy}, {Cesaroni},
  {Garc{\'\i}a-Lario}, {Hoare}, {Huang}, {Joncas}, {Liu}, {Maiolo}, {Marsh},
  {Maruccia}, {M{\`e}ge}, {Peretto}, {Rygl}, {Schilke}, {Thompson},
  {Traficante}, {Umana}, {Veneziani}, {Ward-Thompson}, {Whitworth}, {Arab},
  {Bandieramonte}, {Becciani}, {Brescia}, {Buemi}, {Bufano}, {Butora},
  {Cavuoti}, {Costa}, {Fiorellino}, {Hajnal}, {Hayakawa}, {Kacsuk}, {Leto}, {Li
  Causi}, {Marchili}, {Martinavarro-Armengol}, {Mercurio}, {Molinaro},
  {Riccio}, {Sano}, {Sciacca}, {Tachihara}, {Torii}, {Trigilio}, {Vitello}, \&
  {Yamamoto}}]{Elia2017}
{Elia}, D., {Molinari}, S., {Schisano}, E., {et~al.} 2017, \mnras, 471, 100

\bibitem[{{Frank} {et~al.}(2014){Frank}, {Ray}, {Cabrit}, {Hartigan}, {Arce},
  {Bacciotti}, {Bally}, {Benisty}, {Eisl{\"o}ffel}, {G{\"u}del}, {Lebedev},
  {Nisini}, \& {Raga}}]{2014Frank}
{Frank}, A., {Ray}, T.~P., {Cabrit}, S., {et~al.} 2014, in Protostars and
  Planets VI, ed. H.~{Beuther}, R.~S. {Klessen}, C.~P. {Dullemond}, \&
  T.~{Henning}, 451--474

\bibitem[{{Gerner} {et~al.}(2014){Gerner}, {Beuther}, {Semenov}, {Linz},
  {Vasyunina}, {Bihr}, {Shirley}, \& {Henning}}]{2014Gerner}
{Gerner}, T., {Beuther}, H., {Semenov}, D., {et~al.} 2014, \aap, 563, A97

\bibitem[{{Gieser} {et~al.}(2021){Gieser}, {Beuther}, {Semenov}, {Ahmadi},
  {Suri}, {M{\"o}ller}, {Beltr{\'a}n}, {Klaassen}, {Zhang}, {Urquhart},
  {Henning}, {Feng}, {Galv{\'a}n-Madrid}, {de Souza Magalh{\~a}es},
  {Moscadelli}, {Longmore}, {Leurini}, {Kuiper}, {Peters}, {Menten},
  {Csengeri}, {Fuller}, {Wyrowski}, {Lumsden}, {S{\'a}nchez-Monge}, {Maud},
  {Linz}, {Palau}, {Schilke}, {Pety}, {Pudritz}, {Winters}, \&
  {Pi{\'e}tu}}]{2021Gieser}
{Gieser}, C., {Beuther}, H., {Semenov}, D., {et~al.} 2021, \aap, 648, A66

\bibitem[{{Gieser} {et~al.}(2022){Gieser}, {Beuther}, {Semenov}, {Suri},
  {Soler}, {Linz}, {Syed}, {Henning}, {Feng}, {M{\"o}ller}, {Palau}, {Winters},
  {Beltr{\'a}n}, {Kuiper}, {Moscadelli}, {Klaassen}, {Urquhart}, {Peters},
  {Longmore}, {S{\'a}nchez-Monge}, {Galv{\'a}n-Madrid}, {Pudritz}, \&
  {Johnston}}]{2022Gieser}
{Gieser}, C., {Beuther}, H., {Semenov}, D., {et~al.} 2022, \aap, 657, A3

\bibitem[{{Ginsburg} {et~al.}(2019){Ginsburg}, {Koch}, {Robitaille},
  {Beaumont}, {Adamginsburg}, {ZuHone}, {Sipocz}, {Patra}, {Jones}, {Lim},
  {Rosolowsky}, {Stern}, {Earl}, {De Val-Borro}, {Jrobbfed}, {Shuokong},
  {Kepley}, {Sokolov}, {Badger}, {Maret}, {Garrido}, {Booker}, \&
  {Tollerud}}]{spectralcube}
{Ginsburg}, A., {Koch}, E., {Robitaille}, T., {et~al.} 2019,
  {radio-astro-tools/spectral-cube: v0.4.4}

\bibitem[{{Goldsmith} {et~al.}(2008){Goldsmith}, {Heyer}, {Narayanan}, {Snell},
  {Li}, \& {Brunt}}]{2008Goldsmith}
{Goldsmith}, P.~F., {Heyer}, M., {Narayanan}, G., {et~al.} 2008, \apj, 680, 428

\bibitem[{{G{\'o}mez} \& {V{\'a}zquez-Semadeni}(2014)}]{2014Gomez}
{G{\'o}mez}, G.~C. \& {V{\'a}zquez-Semadeni}, E. 2014, \apj, 791, 124

\bibitem[{{Hacar} {et~al.}(2023){Hacar}, {Clark}, {Heitsch}, {Kainulainen},
  {Panopoulou}, {Seifried}, \& {Smith}}]{2023Hacar}
{Hacar}, A., {Clark}, S.~E., {Heitsch}, F., {et~al.} 2023, in Astronomical
  Society of the Pacific Conference Series, Vol. 534, Protostars and Planets
  VII, ed. S.~{Inutsuka}, Y.~{Aikawa}, T.~{Muto}, K.~{Tomida}, \& M.~{Tamura},
  153

\bibitem[{Harris {et~al.}(2020)Harris, Millman, van~der Walt, Gommers,
  Virtanen, Cournapeau, Wieser, Taylor, Berg, Smith, Kern, Picus, Hoyer, van
  Kerkwijk, Brett, Haldane, del R{\'{i}}o, Wiebe, Peterson,
  G{\'{e}}rard-Marchant, Sheppard, Reddy, Weckesser, Abbasi, Gohlke, \&
  Oliphant}]{numpy}
Harris, C.~R., Millman, K.~J., van~der Walt, S.~J., {et~al.} 2020, Nature, 585,
  357

\bibitem[{{Hartmann} {et~al.}(2012){Hartmann}, {Ballesteros-Paredes}, \&
  {Heitsch}}]{2012Hartmann}
{Hartmann}, L., {Ballesteros-Paredes}, J., \& {Heitsch}, F. 2012, \mnras, 420,
  1457

\bibitem[{{Henshaw} {et~al.}(2014){Henshaw}, {Caselli}, {Fontani},
  {Jim{\'e}nez-Serra}, \& {Tan}}]{2014Henshaw}
{Henshaw}, J.~D., {Caselli}, P., {Fontani}, F., {Jim{\'e}nez-Serra}, I., \&
  {Tan}, J.~C. 2014, \mnras, 440, 2860

\bibitem[{{Hildebrand}(1983)}]{1983H}
{Hildebrand}, R.~H. 1983, \qjras, 24, 267

\bibitem[{{Hoare} {et~al.}(2005){Hoare}, {Lumsden}, {Oudmaijer}, {Urquhart},
  {Busfield}, {Sheret}, {Clarke}, {Moore}, {Allsopp}, {Burton}, {Purcell},
  {Jiang}, \& {Wang}}]{2005Hoare}
{Hoare}, M.~G., {Lumsden}, S.~L., {Oudmaijer}, R.~D., {et~al.} 2005, in Massive
  Star Birth: A Crossroads of Astrophysics, ed. R.~{Cesaroni}, M.~{Felli},
  E.~{Churchwell}, \& M.~{Walmsley}, Vol. 227, 370--375

\bibitem[{Hunter(2007)}]{matplotlib}
Hunter, J.~D. 2007, Computing in Science \& Engineering, 9, 90

\bibitem[{{Izumi} {et~al.}(2024){Izumi}, {Sanhueza}, {Koch}, {Lu}, {Li},
  {Sabatini}, {Olguin}, {Zhang}, {Nakamura}, {Tatematsu}, {Morii}, {Sakai}, \&
  {Tafoya}}]{2024I}
{Izumi}, N., {Sanhueza}, P., {Koch}, P.~M., {et~al.} 2024, \apj, 963, 163

\bibitem[{{Jackson} {et~al.}(2010){Jackson}, {Finn}, {Chambers}, {Rathborne},
  \& {Simon}}]{2010Jackson}
{Jackson}, J.~M., {Finn}, S.~C., {Chambers}, E.~T., {Rathborne}, J.~M., \&
  {Simon}, R. 2010, \apjl, 719, L185

\bibitem[{{Kahn}(1974)}]{1974Kahn}
{Kahn}, F.~D. 1974, \aap, 37, 149

\bibitem[{{Kirk} {et~al.}(2013){Kirk}, {Myers}, {Bourke}, {Gutermuth},
  {Hedden}, \& {Wilson}}]{2013Kirk}
{Kirk}, H., {Myers}, P.~C., {Bourke}, T.~L., {et~al.} 2013, \apj, 766, 115

\bibitem[{{Kuiper} \& {Hosokawa}(2018)}]{2018Kuiper}
{Kuiper}, R. \& {Hosokawa}, T. 2018, \aap, 616, A101

\bibitem[{Lada \& Lada(2003)}]{Lada_2003}
Lada, C.~J. \& Lada, E.~A. 2003, Annual Review of Astronomy and Astrophysics,
  41, 57–115

\bibitem[{{L{\'o}pez-Sepulcre} {et~al.}(2010){L{\'o}pez-Sepulcre}, {Cesaroni},
  \& {Walmsley}}]{2010Lopez}
{L{\'o}pez-Sepulcre}, A., {Cesaroni}, R., \& {Walmsley}, C.~M. 2010, \aap, 517,
  A66

\bibitem[{{Lumsden} {et~al.}(2013){Lumsden}, {Hoare}, {Urquhart}, {Oudmaijer},
  {Davies}, {Mottram}, {Cooper}, \& {Moore}}]{lumsden2013}
{Lumsden}, S.~L., {Hoare}, M.~G., {Urquhart}, J.~S., {et~al.} 2013, \apjs, 208,
  11

\bibitem[{{Mangum} \& {Wootten}(1993)}]{mangum1993}
{Mangum}, J.~G. \& {Wootten}, A. 1993, \apjs, 89, 123

\bibitem[{{McKee} \& {Tan}(2003)}]{2003McKee}
{McKee}, C.~F. \& {Tan}, J.~C. 2003, \apj, 585, 850

\bibitem[{McKnight \& Najab(2010)}]{mcknight2010mann}
McKnight, P.~E. \& Najab, J. 2010, The Corsini encyclopedia of psychology, 1

\bibitem[{{M{\`e}ge} {et~al.}(2021){M{\`e}ge}, {Russeil}, {Zavagno}, {Elia},
  {Molinari}, {Brunt}, {Butora}, {Cambresy}, {Di Giorgio}, {Fenouillet},
  {Fukui}, {Lambert}, {Makai}, {Merello}, {Meunier}, {Molinaro}, {Moreau},
  {Pezzuto}, {Poulin}, {Schisano}, \& {Schuller}}]{2021Mege}
{M{\`e}ge}, P., {Russeil}, D., {Zavagno}, A., {et~al.} 2021, \aap, 646, A74

\bibitem[{{Molinari} {et~al.}(2019){Molinari}, {Baldeschi}, {Robitaille},
  {Morales}, {Schisano}, {Traficante}, {Merello}, {Molinaro}, {Vitello},
  {Sciacca}, \& {Liu}}]{2019Molinari}
{Molinari}, S., {Baldeschi}, A., {Robitaille}, T.~P., {et~al.} 2019, \mnras,
  486, 4508

\bibitem[{{Molinari} {et~al.}(2016){Molinari}, {Merello}, {Elia}, {Cesaroni},
  {Testi}, \& {Robitaille}}]{molinari2016}
{Molinari}, S., {Merello}, M., {Elia}, D., {et~al.} 2016, \apjl, 826, L8

\bibitem[{{Molinari} {et~al.}(2010{\natexlab{a}}){Molinari}, {Swinyard},
  {Bally}, {Barlow}, {Bernard}, {Martin}, {Moore}, {Noriega-Crespo}, {Plume},
  {Testi}, {Zavagno}, {Abergel}, {Ali}, {Anderson}, {Andr{\'e}}, {Baluteau},
  {Battersby}, {Beltr{\'a}n}, {Benedettini}, {Billot}, {Blommaert}, {Bontemps},
  {Boulanger}, {Brand}, {Brunt}, {Burton}, {Calzoletti}, {Carey}, {Caselli},
  {Cesaroni}, {Cernicharo}, {Chakrabarti}, {Chrysostomou}, {Cohen},
  {Compiegne}, {de Bernardis}, {de Gasperis}, {di Giorgio}, {Elia}, {Faustini},
  {Flagey}, {Fukui}, {Fuller}, {Ganga}, {Garcia-Lario}, {Glenn}, {Goldsmith},
  {Griffin}, {Hoare}, {Huang}, {Ikhenaode}, {Joblin}, {Joncas}, {Juvela},
  {Kirk}, {Lagache}, {Li}, {Lim}, {Lord}, {Marengo}, {Marshall}, {Masi},
  {Massi}, {Matsuura}, {Minier}, {Miville-Desch{\^e}nes}, {Montier}, {Morgan},
  {Motte}, {Mottram}, {M{\"u}ller}, {Natoli}, {Neves}, {Olmi}, {Paladini},
  {Paradis}, {Parsons}, {Peretto}, {Pestalozzi}, {Pezzuto}, {Piacentini},
  {Piazzo}, {Polychroni}, {Pomar{\`e}s}, {Popescu}, {Reach}, {Ristorcelli},
  {Robitaille}, {Robitaille}, {Rod{\'o}n}, {Roy}, {Royer}, {Russeil},
  {Saraceno}, {Sauvage}, {Schilke}, {Schisano}, {Schneider}, {Schuller},
  {Schulz}, {Sibthorpe}, {Smith}, {Smith}, {Spinoglio}, {Stamatellos},
  {Strafella}, {Stringfellow}, {Sturm}, {Taylor}, {Thompson}, {Traficante},
  {Tuffs}, {Umana}, {Valenziano}, {Vavrek}, {Veneziani}, {Viti}, {Waelkens},
  {Ward-Thompson}, {White}, {Wilcock}, {Wyrowski}, {Yorke}, \&
  {Zhang}}]{molinari2010}
{Molinari}, S., {Swinyard}, B., {Bally}, J., {et~al.} 2010{\natexlab{a}}, \aap,
  518, L100

\bibitem[{{Molinari} {et~al.}(2010{\natexlab{b}}){Molinari}, {Swinyard},
  {Bally}, {Barlow}, {Bernard}, {Martin}, {Moore}, {Noriega-Crespo}, {Plume},
  {Testi}, {Zavagno}, {Abergel}, {Ali}, {Andr{\'e}}, {Baluteau}, {Benedettini},
  {Bern{\'e}}, {Billot}, {Blommaert}, {Bontemps}, {Boulanger}, {Brand},
  {Brunt}, {Burton}, {Campeggio}, {Carey}, {Caselli}, {Cesaroni}, {Cernicharo},
  {Chakrabarti}, {Chrysostomou}, {Codella}, {Cohen}, {Compiegne}, {Davis}, {de
  Bernardis}, {de Gasperis}, {Di Francesco}, {di Giorgio}, {Elia}, {Faustini},
  {Fischera}, {Fukui}, {Fuller}, {Ganga}, {Garcia-Lario}, {Giard}, {Giardino},
  {Glenn}, {Goldsmith}, {Griffin}, {Hoare}, {Huang}, {Jiang}, {Joblin},
  {Joncas}, {Juvela}, {Kirk}, {Lagache}, {Li}, {Lim}, {Lord}, {Lucas},
  {Maiolo}, {Marengo}, {Marshall}, {Masi}, {Massi}, {Matsuura}, {Meny},
  {Minier}, {Miville-Desch{\^e}nes}, {Montier}, {Motte}, {M{\"u}ller},
  {Natoli}, {Neves}, {Olmi}, {Paladini}, {Paradis}, {Pestalozzi}, {Pezzuto},
  {Piacentini}, {Pomar{\`e}s}, {Popescu}, {Reach}, {Richer}, {Ristorcelli},
  {Roy}, {Royer}, {Russeil}, {Saraceno}, {Sauvage}, {Schilke},
  {Schneider-Bontemps}, {Schuller}, {Schultz}, {Shepherd}, {Sibthorpe},
  {Smith}, {Smith}, {Spinoglio}, {Stamatellos}, {Strafella}, {Stringfellow},
  {Sturm}, {Taylor}, {Thompson}, {Tuffs}, {Umana}, {Valenziano}, {Vavrek},
  {Viti}, {Waelkens}, {Ward-Thompson}, {White}, {Wyrowski}, {Yorke}, \&
  {Zhang}}]{2010Molinari}
{Molinari}, S., {Swinyard}, B., {Bally}, J., {et~al.} 2010{\natexlab{b}},
  \pasp, 122, 314

\bibitem[{{Motte} {et~al.}(2018){Motte}, {Bontemps}, \& {Louvet}}]{2018Motte}
{Motte}, F., {Bontemps}, S., \& {Louvet}, F. 2018, \araa, 56, 41

\bibitem[{Motte {et~al.}(2018)Motte, Bontemps, \& Louvet}]{Motte_2018}
Motte, F., Bontemps, S., \& Louvet, F. 2018, Annual Review of Astronomy and
  Astrophysics, 56, 41–82

\bibitem[{{Myers}(2009)}]{2009Myers}
{Myers}, P.~C. 2009, \apj, 700, 1609

\bibitem[{{Offner} {et~al.}(2014){Offner}, {Clark}, {Hennebelle}, {Bastian},
  {Bate}, {Hopkins}, {Moraux}, \& {Whitworth}}]{2014Offner}
{Offner}, S.~S.~R., {Clark}, P.~C., {Hennebelle}, P., {et~al.} 2014, in
  Protostars and Planets VI, ed. H.~{Beuther}, R.~S. {Klessen}, C.~P.
  {Dullemond}, \& T.~{Henning}, 53--75

\bibitem[{{Olguin} {et~al.}(2023){Olguin}, {Sanhueza}, {Chen}, {Lu}, {Oya},
  {Zhang}, {Ginsburg}, {Taniguchi}, {Li}, {Morii}, {Sakai}, \&
  {Nakamura}}]{2023O}
{Olguin}, F.~A., {Sanhueza}, P., {Chen}, H.-R.~V., {et~al.} 2023, \apjl, 959,
  L31

\bibitem[{{Ossenkopf} \& {Henning}(1994)}]{ossenkopf1994}
{Ossenkopf}, V. \& {Henning}, T. 1994, \aap, 291, 943

\bibitem[{{Padoan} {et~al.}(2020{\natexlab{a}}){Padoan}, {Pan}, {Juvela},
  {Haugb{\o}lle}, \& {Nordlund}}]{2020Padoan}
{Padoan}, P., {Pan}, L., {Juvela}, M., {Haugb{\o}lle}, T., \& {Nordlund},
  {\r{A}}. 2020{\natexlab{a}}, \apj, 900, 82

\bibitem[{{Padoan} {et~al.}(2020{\natexlab{b}}){Padoan}, {Pan}, {Juvela},
  {Haugb{\o}lle}, \& {Nordlund}}]{Padoan2020}
{Padoan}, P., {Pan}, L., {Juvela}, M., {Haugb{\o}lle}, T., \& {Nordlund},
  {\r{A}}. 2020{\natexlab{b}}, \apj, 900, 82

\bibitem[{{Peretto} {et~al.}(2013){Peretto}, {Fuller}, {Duarte-Cabral},
  {Avison}, {Hennebelle}, {Pineda}, {Andr{\'e}}, {Bontemps}, {Motte},
  {Schneider}, \& {Molinari}}]{2013Peretto}
{Peretto}, N., {Fuller}, G.~A., {Duarte-Cabral}, A., {et~al.} 2013, \aap, 555,
  A112

\bibitem[{{Redaelli} {et~al.}(2022){Redaelli}, {Bovino}, {Sanhueza}, {Morii},
  {Sabatini}, {Caselli}, {Giannetti}, \& {Li}}]{2022Redaelli}
{Redaelli}, E., {Bovino}, S., {Sanhueza}, P., {et~al.} 2022, \apj, 936, 169

\bibitem[{{Salpeter}(1955)}]{Salpeter}
{Salpeter}, E.~E. 1955, \apj, 121, 161

\bibitem[{{Sanhueza} {et~al.}(2021){Sanhueza}, {Girart}, {Padovani}, {Galli},
  {Hull}, {Zhang}, {Cortes}, {Stephens}, {Fern{\'a}ndez-L{\'o}pez}, {Jackson},
  {Frau}, {Kock}, {Wu}, {Zapata}, {Olguin}, {Lu}, {Silva}, {Tang}, {Sakai},
  {Guzm{\'a}n}, {Tatematsu}, {Nakamura}, \& {Chen}}]{2021Sanhueza}
{Sanhueza}, P., {Girart}, J.~M., {Padovani}, M., {et~al.} 2021, \apjl, 915, L10

\bibitem[{{Scalo}(1985)}]{1985Scalo}
{Scalo}, J.~M. 1985, in Protostars and Planets II, ed. D.~C. {Black} \& M.~S.
  {Matthews}, 201--296

\bibitem[{{Schisano} {et~al.}(2020){Schisano}, {Molinari}, {Elia},
  {Benedettini}, {Olmi}, {Pezzuto}, {Traficante}, {Brescia}, {Cavuoti}, {di
  Giorgio}, {Liu}, {Moore}, {Noriega-Crespo}, {Riccio}, {Baldeschi},
  {Becciani}, {Peretto}, {Merello}, {Vitello}, {Zavagno}, {Beltr{\'a}n},
  {Cambr{\'e}sy}, {Eden}, {Li Causi}, {Molinaro}, {Palmeirim}, {Sciacca},
  {Testi}, {Umana}, \& {Whitworth}}]{2020Schisano}
{Schisano}, E., {Molinari}, S., {Elia}, D., {et~al.} 2020, \mnras, 492, 5420

\bibitem[{{Schneider} {et~al.}(2010){Schneider}, {Csengeri}, {Bontemps},
  {Motte}, {Simon}, {Hennebelle}, {Federrath}, \& {Klessen}}]{2010Schneider}
{Schneider}, N., {Csengeri}, T., {Bontemps}, S., {et~al.} 2010, \aap, 520, A49

\bibitem[{{Schuller} {et~al.}(2009){Schuller}, {Menten}, {Contreras},
  {Wyrowski}, {Schilke}, {Bronfman}, {Henning}, {Walmsley}, {Beuther},
  {Bontemps}, {Cesaroni}, {Deharveng}, {Garay}, {Herpin}, {Lefloch}, {Linz},
  {Mardones}, {Minier}, {Molinari}, {Motte}, {Nyman}, {Reveret}, {Risacher},
  {Russeil}, {Schneider}, {Testi}, {Troost}, {Vasyunina}, {Wienen}, {Zavagno},
  {Kovacs}, {Kreysa}, {Siringo}, \& {Wei{\ss}}}]{2009Schuller}
{Schuller}, F., {Menten}, K.~M., {Contreras}, Y., {et~al.} 2009, \aap, 504, 415

\bibitem[{{Shirley}(2015)}]{shirley2015}
{Shirley}, Y.~L. 2015, \pasp, 127, 299

\bibitem[{{Smith} {et~al.}(2009){Smith}, {Longmore}, \& {Bonnell}}]{2009Smith}
{Smith}, R.~J., {Longmore}, S., \& {Bonnell}, I. 2009, \mnras, 400, 1775

\bibitem[{{Svoboda} {et~al.}(2019){Svoboda}, {Shirley}, {Traficante},
  {Battersby}, {Fuller}, {Zhang}, {Beuther}, {Peretto}, {Brogan}, \&
  {Hunter}}]{2019Svoboda}
{Svoboda}, B.~E., {Shirley}, Y.~L., {Traficante}, A., {et~al.} 2019, \apj, 886,
  36

\bibitem[{{Syed} {et~al.}(2022){Syed}, {Soler}, {Beuther}, {Wang}, {Suri},
  {Henshaw}, {Riener}, {Bialy}, {Rezaei Kh.}, {Stil}, {Goldsmith}, {Rugel},
  {Glover}, {Klessen}, {Kerp}, {Urquhart}, {Ott}, {Roy}, {Schneider}, {Smith},
  {Longmore}, \& {Linz}}]{2022Syed}
{Syed}, J., {Soler}, J.~D., {Beuther}, H., {et~al.} 2022, \aap, 657, A1

\bibitem[{{Tackenberg} {et~al.}(2014){Tackenberg}, {Beuther}, {Henning},
  {Linz}, {Sakai}, {Ragan}, {Krause}, {Nielbock}, {Hennemann}, {Pitann}, \&
  {Schmiedeke}}]{2014Tackenberg}
{Tackenberg}, J., {Beuther}, H., {Henning}, T., {et~al.} 2014, \aap, 565, A101

\bibitem[{{Tan} {et~al.}(2014){Tan}, {Beltr{\'a}n}, {Caselli}, {Fontani},
  {Fuente}, {Krumholz}, {McKee}, \& {Stolte}}]{2014T}
{Tan}, J.~C., {Beltr{\'a}n}, M.~T., {Caselli}, P., {et~al.} 2014, 149

\bibitem[{{Thomasson} {et~al.}(2022){Thomasson}, {Joncour}, {Moraux},
  {Crespelle}, {Motte}, {Pouteau}, \& {Nony}}]{2022Thomasson}
{Thomasson}, B., {Joncour}, I., {Moraux}, E., {et~al.} 2022, \aap, 665, A119

\bibitem[{{Traficante} {et~al.}(2017){Traficante}, {Fuller}, {Billot},
  {Duarte-Cabral}, {Merello}, {Molinari}, {Peretto}, \&
  {Schisano}}]{2017Traficante}
{Traficante}, A., {Fuller}, G.~A., {Billot}, N., {et~al.} 2017, \mnras, 470,
  3882

\bibitem[{{Traficante} {et~al.}(2023){Traficante}, {Jones}, {Avison}, {Fuller},
  {Benedettini}, {Elia}, {Molinari}, {Peretto}, {Pezzuto}, {Pillai}, {Rygl},
  {Schisano}, \& {Smith}}]{traficante2023}
{Traficante}, A., {Jones}, B.~M., {Avison}, A., {et~al.} 2023, \mnras, 520,
  2306

\bibitem[{{Urquhart} {et~al.}(2007){Urquhart}, {Busfield}, {Hoare}, {Lumsden},
  {Oudmaijer}, {Moore}, {Gibb}, {Purcell}, {Burton}, \&
  {Marechal}}]{2007Urquhart}
{Urquhart}, J.~S., {Busfield}, A.~L., {Hoare}, M.~G., {et~al.} 2007, \aap, 474,
  891

\bibitem[{{Urquhart} {et~al.}(2018){Urquhart}, {K{\"o}nig}, {Giannetti},
  {Leurini}, {Moore}, {Eden}, {Pillai}, {Thompson}, {Braiding}, {Burton},
  {Csengeri}, {Dempsey}, {Figura}, {Froebrich}, {Menten}, {Schuller}, {Smith},
  \& {Wyrowski}}]{2018Urquhart}
{Urquhart}, J.~S., {K{\"o}nig}, C., {Giannetti}, A., {et~al.} 2018, \mnras,
  473, 1059

\bibitem[{{Urquhart} {et~al.}(2022){Urquhart}, {Wells}, {Pillai}, {Leurini},
  {Giannetti}, {Moore}, {Thompson}, {Figura}, {Colombo}, {Yang}, {K{\"o}nig},
  {Wyrowski}, {Menten}, {Rigby}, {Eden}, \& {Ragan}}]{2022Urquhart}
{Urquhart}, J.~S., {Wells}, M.~R.~A., {Pillai}, T., {et~al.} 2022, \mnras, 510,
  3389

\bibitem[{{van der Tak} {et~al.}(2007){van der Tak}, {Black}, {Sch{\"o}ier},
  {Jansen}, \& {van Dishoeck}}]{2007tak}
{van der Tak}, F.~F.~S., {Black}, J.~H., {Sch{\"o}ier}, F.~L., {Jansen}, D.~J.,
  \& {van Dishoeck}, E.~F. 2007, \aap, 468, 627

\bibitem[{{van Gelder} {et~al.}(2021){van Gelder}, {Tabone}, {van Dishoeck}, \&
  {Godard}}]{2021VG}
{van Gelder}, M.~L., {Tabone}, B., {van Dishoeck}, E.~F., \& {Godard}, B. 2021,
  \aap, 653, A159

\bibitem[{{Widmann} {et~al.}(2016){Widmann}, {Beuther}, {Schilke}, \&
  {Stanke}}]{2016W}
{Widmann}, F., {Beuther}, H., {Schilke}, P., \& {Stanke}, T. 2016, \aap, 589,
  A29

\bibitem[{{Williams} {et~al.}(2000){Williams}, {Blitz}, \&
  {McKee}}]{2000Williams}
{Williams}, J.~P., {Blitz}, L., \& {McKee}, C.~F. 2000, in Protostars and
  Planets IV, ed. V.~{Mannings}, A.~P. {Boss}, \& S.~S. {Russell}, 97

\bibitem[{{Wolfire} \& {Cassinelli}(1987)}]{1987Wolfire}
{Wolfire}, M.~G. \& {Cassinelli}, J.~P. 1987, \apj, 319, 850

\bibitem[{{Yorke} \& {Kruegel}(1977)}]{1977Yorke}
{Yorke}, H.~W. \& {Kruegel}, E. 1977, \aap, 54, 183

\bibitem[{{Zhang} {et~al.}(2015){Zhang}, {Wang}, {Lu}, \&
  {Jim{\'e}nez-Serra}}]{2015Zhang}
{Zhang}, Q., {Wang}, K., {Lu}, X., \& {Jim{\'e}nez-Serra}, I. 2015, \apj, 804,
  141

\bibitem[{{Zinnecker}(1984)}]{1984MZinnecker}
{Zinnecker}, H. 1984, \mnras, 210, 43

\bibitem[{{Zinnecker} \& {Yorke}(2007)}]{2007ZY}
{Zinnecker}, H. \& {Yorke}, H.~W. 2007, \araa, 45, 481

\end{thebibliography}

\begin{appendix}
\onecolumn
\section{Sample Parameter Histograms}
\label{sect:appendixa}
\begin{figure*}[ht!]
    \centering
    \includegraphics[width=0.90\textwidth, height=0.52\textheight]{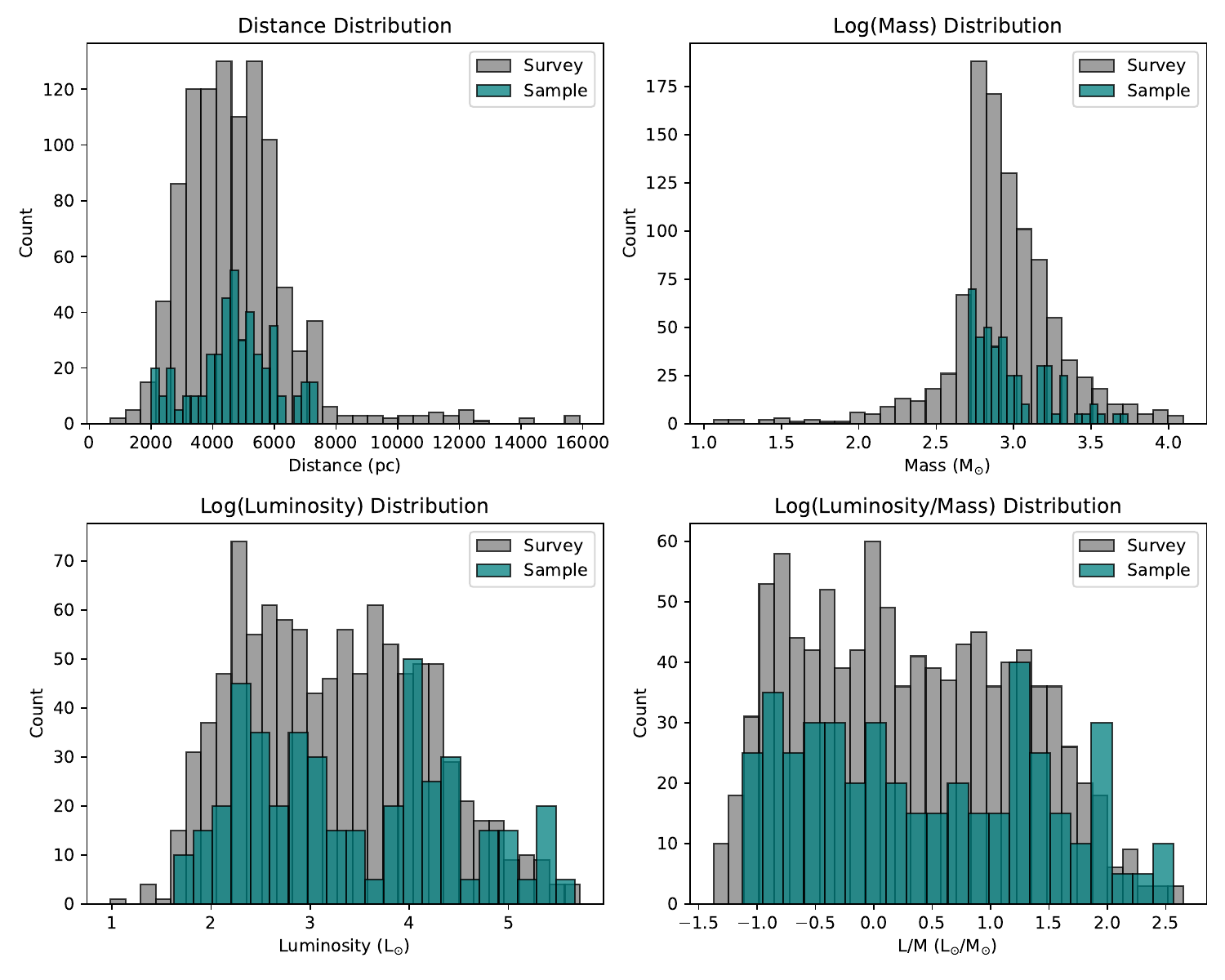}
    \caption{Histograms comparing the all regions in the ALMAGAL survey vs the ones chosen for this sample for distance, mass, luminosity, and L/M.}
    \label{fig:surveysample}
\end{figure*}

\section{Derivation of flow rate along a filament}
\label{sect:appendixb}
In this appendix chapter, we present a derivation of the formula used for the observational measurement of the flow rates along filamentary structures.
We start from mass conservation in hydrodynamics, namely the continuity equation
\begin{equation}
    \frac{\partial}{\partial t} \rho 
    + 
    \nabla \cdot \left( \rho \vec{v} \right) 
    = 
    0
\end{equation}
with the density $\rho$ and the velocity $\vec{v}$. If we represent the filament as a cylindrical object, and we check for the temporal change of mass within a section of the cylinder of real length $w_r$ and fixed volume $V$, we can take the volume out of the time derivative:
\begin{equation}
    \frac{\partial}{\partial t} M_V 
    =
    \dot{M}_V
    =
    - V \nabla \cdot \left( \rho \vec{v} \right) 
\end{equation}
If we further approximate the medium density to be uniform along the spatial scale of interest $w_r$, we take the density out of the spatial derivative, and the volume $V$ cancels out:
\begin{equation}
    \dot{M}_V
    =
    - M_V \nabla \cdot \vec{v}
\end{equation}
For a one-dimensional flow along the filament, we can approximate the divergence of the velocity field $\nabla \cdot \vec{v}$ as the velocity difference $\Delta v_\mathrm{out-in}$ of the flow out of the section of length $w_r$ and into it:
\begin{equation}
\label{eq:mass_conservation}
    \dot{M}_V
    =
    - M_V \frac{\Delta v_\mathrm{out-in}}{w_r}
\end{equation}
Eq.~\eqref{eq:mass_conservation} represents the mass conservation within the section $w_r$ of the filament.
As an example: For a uniform velocity field, the inflow and outflow velocities are identical $\Delta v_\mathrm{out-in} = 0$ and the mass within the section remains unchanged.

To obtain a formula for the flow rate $\dot{M}_r$, we have to substitute the velocity difference by the absolute, local velocity $v_r$ of the flow (subscript "r" meaning the real values and subscript "obs" the corresponding observed values):
\begin{equation}
    \dot{M}_r
    =
    M_V \frac{|v_r|}{w_r}
\end{equation}
Here, we included the convention that the flow rate is always treated as a positive value, regardless if the flow is pointing towards the observer or away from the observer.

So far, we have discussed the system only in its local (unobserved) properties.
Now let's introduce the observational properties:
The absolute, local velocity $v_r$ can be obtained from the observational data by subtracting the systemic velocity $v_\mathrm{sys}$ from the observed velocity $v_\mathrm{obs}$ with $\Delta v_\mathrm{r} = |v_\mathrm{obs} - v_\mathrm{sys}|$:
\begin{equation}
\label{eq:flow_rate}
    \dot{M}_r
    =
    M_V \frac{\Delta v_\mathrm{r}}{w_r}.
\end{equation}
The mass $M_V$ can be approximated by the column density $\Sigma$ times the beam. Since we use 1$''$ length scale, roughly a beam width, we can approximate the beam with
\begin{equation}
  A_\mathrm{beam} \sim  w_\mathrm{obs}^{2}. 
 \end{equation}
 Substituting in the beam area we get
 \begin{equation}
 \label{eq:param1}
  M_V  =  \Sigma_{\mathrm{obs}} \cdot A_\mathrm{beam} = \Sigma_{\mathrm{obs}} \cdot w_\mathrm{obs}^{2},
\end{equation}
where we chose our measured length scale as the size of the beam. 
Including now the inclination dependence of the observed parameters:
\begin{equation}
\label{eq:param2}
    \Delta v_\mathrm{r} = \frac{\Delta v_\mathrm{obs}}{\sin{i}}
\end{equation}
and 
\begin{equation}
\label{eq:param3}
     w_\mathrm{r} = \frac{w_\mathrm{obs}}{\cos{i}}.
\end{equation}
Substituting Equations~\eqref{eq:param1}-\eqref{eq:param3} into Equation~\eqref{eq:flow_rate} we get
\begin{equation}
     \dot{M}_r = \Sigma_{\mathrm{obs}} \cdot w_\mathrm{obs}^{2} \cdot \frac{\Delta v_\mathrm{obs}}{w_\mathrm{obs}} \cdot \frac{\cos{i}}{\sin{i}},
\end{equation}
which in its final form is 
\begin{equation}
      \dot{M}_r = \Sigma_{\mathrm{obs}}  \cdot \frac{\Delta v_{\mathrm{obs}}}{\tan(i)} \cdot w_{\mathrm{obs}}.
\end{equation}

\vspace{8cm}
\pagebreak
\section{Core Mass Figures}
\label{sect:appendixc}
\begin{figure*}[h]
\begin{subfigure}{0.5\textwidth}
\includegraphics[width=0.99\textwidth, height=0.28\textheight]{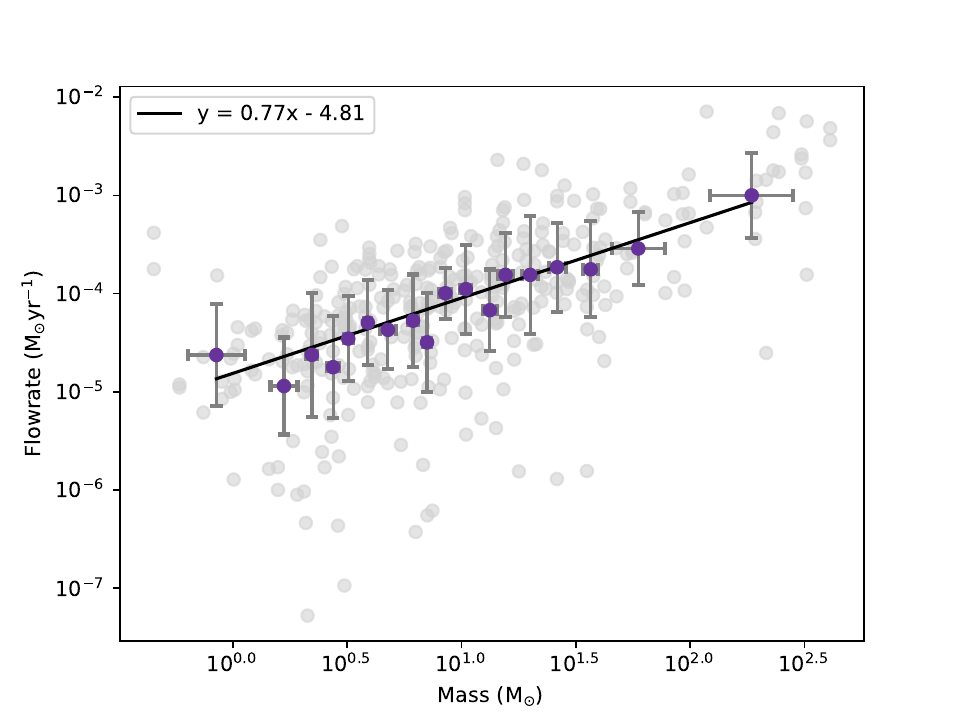} 
\caption{Only inner flow rates.}
\label{fig:inner}
\end{subfigure}
\begin{subfigure}{0.5\textwidth}
\includegraphics[width=0.99\textwidth, height=0.28\textheight]{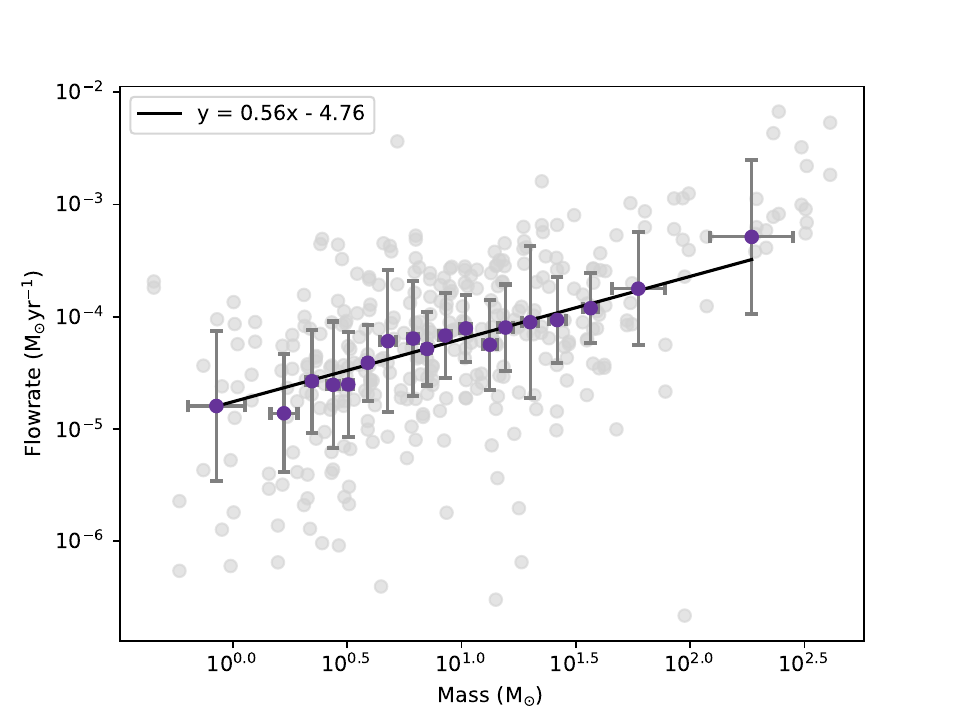}
\caption{Only outer flow rates.}
\label{fig:outer}
\end{subfigure}
\caption{Flow rate vs. core mass relation for only the flow rates closer to the core (panel (a)), and only the flow rates further from the core (panel (b)).}
\end{figure*}

\section{Source and Core Parameters}
\label{sect:appendixd}
\begin{longtable}{llllllllll}
\caption{\label{sources} Table of source parameters (10 row preview).}\\
\hline\hline
ALMAGAL Name & ID & $ G_{\mathrm{lon}}$ & $G_{\mathrm{lat}}$ & $v_{\mathrm{lsr}}$ & Distance & Mass & Luminosity & Temp & Classification\\
 &  & (deg) & (deg) & (kms$^-1$) & (pc) & (M$_{\sun}$) & (L$_{\sun}$) & (K) & \\
\hline\hline
\endfirsthead
\caption{continued.}\\
\hline\hline
ALMAGAL Name & ID & $G_{\mathrm{lon}}$ & $G_{\mathrm{lat}}$ & $v_{\mathrm{lsr}}$ & Distance & Mass & Luminosity & Temp & Classification\\
 &  & (deg) & (deg) & (kms$^-1$) & (pc) & (M$_{\sun}$) & (L$_{\sun}$) & (K) & \\
\hline\hline
\endhead
\hline
\endfoot

\begin{filecontents*}{source_table_edited_short.csv}
AG_DESIGNATION,ID,GLON,GLAT,VLSR_AG,DIST_AG,MASS_AG,LUM_AG,MODEL TEMP,CLASSIFICATION
AG022.5316-0.1923,99331,22.533,-0.191,74.8,4470,1503,217,20,Quiescent
AG024.0046+0.0397,107003,24.006,0.040,113.1,5960,2025,691,20,Quiescent
AG024.0147+0.0487,107032,24.014,0.048,111.5,5890,3263,2155,20,Quiescent
AG024.8555+0.0051,111167,24.855,0.005,108.7,5780,806,231,20,Quiescent
AG026.6280-0.0647,119601,26.628,-0.064,103.2,5570,620,279,20,Quiescent
AG027.7975+0.1501,124229,27.798,0.150,81.7,4610,927,108,20,Quiescent
AG028.3550+0.0728,126186,28.354,0.070,81.4,4590,1126,179,20,Quiescent
AG030.2747-0.2311,135558,30.274,-0.231,103.6,5800,1238,243,20,Quiescent
AG031.0225-0.1113,139543,31.022,-0.111,77.1,4400,782,114,20,Quiescent
AG308.8759+0.1730,717276,308.876,0.174,-49.8,4110,557,80,20,Quiescent
\end{filecontents*}
\csvreader{source_table_edited_short.csv}{}{\csvcoli & \csvcolii &\csvcoliii & \csvcoliv & \csvcolv & \csvcolvi & \csvcolvii & \csvcolviii & \csvcolix & \csvcolx \\}
\end{longtable}

\begin{longtable}{llllllll}
\caption{\label{core} Table of core parameters (10 row preview).}\\
\hline\hline
ID & x Offset (") & y Offset (") & Flow rate (L$_\mathrm{outer}$)  &  Flow rate (L$_\mathrm{inner}$) & Flow rate (R$_\mathrm{inner}$) & Flow rate (R$_\mathrm{outer}$) & Mass\\
 &  &  & (\flowrate) & (\flowrate) & (\flowrate) & (\flowrate) & (M$_{\sun}$)\\
\hline\hline

\begin{filecontents*}{pub_cores_edited_short.csv}
object,x_offset,y_offset,flow_left_out,flow_left_in,flow_right_in,flow_right_out,mass_estimate,_idx
99331,2.73,2.52,0.21,0.49,0.02,0.02,17.8,0
107003,2.00,3.99,0.08,0.004,0.23,0.19,6.3,2
107032,0.95,-6.18,1.22,10.27,0.97,0.38,37.6,0
111167,1.63,-2.40,2.84,0.54,0.50,0.003,14.1,0
119601,-0.38,2.00,4.87,0.31,0.40,0.89,6.3,0
124229,1.05,0.00,0.90,0.80,1.57,0.44,6.6,0
126186,4.83,-6.72,0.35,0.38,0.06,0.15,2.7,0
126186,5.04,-3.36,0.20,0.27,0.08,0.12,3.9,1
126186,-3.78,-0.21,0.81,2.46,0.71,1.13,17.8,2
126186,0.42,3.57,2.23,2.65,0.13,0.08,8.4,3
\end{filecontents*}
\csvreader{pub_cores_edited_short.csv}{}{\csvcoli & \csvcolii &\csvcoliii & \csvcoliv & \csvcolv & \csvcolvi & \csvcolvii & \csvcolviii\\}\\\hline\\

\end{longtable}
\end{appendix}
\end{document}